# Transparent Synchronous Dataflow

Steven W.T. Cheung[a], Dan R. Ghica[a], and Koko Muroya[a,b]

a University of Birmingham

b RIMS, Kyoto University

**Abstract** Dataflow programming is a popular and convenient programming paradigm in systems modelling, optimisation, and machine learning. It has a number of advantages, for instance the lacks of control flow allows computation to be carried out in parallel as well as in distributed machines. More recently the idea of dataflow graphs has also been brought into the design of various deep learning frameworks. They facilitate an easy and efficient implementation of automatic differentiation, which is the heart of modern deep learning paradigm.

Many dataflow languages are 'modal' in the sense that the dataflow graph is 'constructed' in an ambient environment then 'executed' using special commands. Simulink, earlier Tensorflow, Lucid Synchrone (LS) or Self-adjusting computation (SAC) are examples of such languages. We are interested in modal dataflow languages in which the ambient environment is a fully fledged functional and imperative language. Such ambient languages make creating complex and large models easier, but they raise questions of language design, safety and efficiency.

LS provides a way to define dataflow graphs through co-recursive equations of *streams*. In these cases the interesting issue is efficiency, particularly of memory utilisation. A semantically interesting question occurs in the case of languages which allow the explicit manipulation of the dataflow graph using imperative constructs. This is the case of early Tensorflow and imperative SAC. However the meta-programming style of TensorFlow in which special commands are used to construct imperatively a dataflow graph was considered awkward and error prone, and TensorFlow moved away from it. Constructing a dataflow graph imperatively can be convenient, but both Tensorflow and SAC are unsafe, in the sense that 'illegal' dataflow graphs can be constructed during the execution, which leads to unsafe behaviour. These problems can be avoided by a judicious language design.

We present an idealised calculus for dataflow with both imperative and functional features using an abstract machine model based on Geometry of Interaction enhanced with rewriting features ('Dynamic GoI'). Although the language is *modal*, with distinct executions for the PCF-like fragment and the DFG fragment, the syntax of the language is uniform. So the language lacks the "metaprogramming" feel of TensorFlow and other embedded DSL-like solutions. Concretely, any operator is used in the same way whether as a part of the host language or the DFG. This is akin to operator overloading, except that it is realised an a purely semantic way. Establishing a "semantic context", defined by its history of execution, is as far as we know a novel approach which, as we shall see, will require a novel approach to semantics.

We have also proved safety of the language and its in-principle efficiency. A prototype implementation of the calculus is also described, as a PPX extension of the OCaml programming language.

# 1 Introduction

## 1.1 Background

Dataflow programming is a broad programming paradigm loosely based on the principle that computation is organised as a graph in which data-changes to the inputs are propagated to the outputs in a more-or-less automatic manner. Notable examples include spreadsheets [51], system analysis and simulation [16], machine learning [1], embedded reactive systems [28], functional reactive programming (FRP) [18] and more.

One of the attractive features of dataflow systems is that their native mode of operation lacks control flow, so it is inherently parallel. This makes issues such as synchrony and (non)determinism pertinent. The parallelism of dataflow system is transparent, with much of the event-level manipulation hidden from the programmer, so 'lower level' issues such as deadlock are usually not a concern. Pure dataflow programming is convenient, but only suitable for particular applications. So it is more common to have dataflow-style idioms and libraries within general purpose programming. This is the case with FRP. The language design challenge is non-trivial. When the host language is used natively to construct the dataflow graph gross inefficiency can be a serious problem, particularly when streams of data must be *merged*, which leads to explosive memory requirements. This is unavoidable in monadic FRP, where the monad multiplication requires the merging of a stream of streams into a single stream. Different languages deal with this problems in different ways, for example by replacing monads with 'arrows' in FRP [37], by using temporal type systems [33], or by clocking each data stream in such a way that streams can be merged using fixed-size buffers [12].

Whereas dataflow functionality can be seamlessly integrated into a larger language (FRP for Haskell) 'modal' behaviour, consisting of a model-building mode and a model-execution mode remains popular for nich languages. Simulink [16], earlier Tensorflow [1], or Lucid Synchrone [40] are examples of such languages. These languages have two distinct modes of execution, one that constructs the dataflow model using a more expressive host language, and one that executes the dataflow model. An extremal example are spreadsheet programs in which the model-building mode is usually the manual manipulation of the spreadsheet itself in the application, and the model execution mode is the fully transparent propagation of changes whenever the value of a spreadsheet cell is changed.

More recently, these dataflow oriented programming languages have become attractive because they facilitate an easy and efficient implementation of *automatic differentiation* (AD) [27] . Most conventional AD techniques apply to so-called "straight line code" which corresponds to ground-type recursion-free pure functional code. Because models in general have type $\mathbb{R}^n \to \mathbb{R}^n$ the restriction to straight-line code is not crippling. However, in practice constructing large models of a complex graph topology, for instance a neural network with tens of thousands of parameters, requires that the model is created using a more expressive programming language. Only the runtime version of the model is then straight-line code. There are some notable exceptions to the rule, for example LSTM neural networks which have a stateful model [22], but this is a very popular programming paradigm. The straight-line-code model can be constructed in meta-programming style in platforms such as early TensorFlow, but this programming model is considered awkward, although very suitable for AD. Models specified seamlessly in an expressive programming languages are a more convenient way to program, but interaction with AD becomes much more difficult. Applying AD to higher order languages with or without effects remains an active area of research without definitive solutions [6].

### 1.2 Problem statement and contribution

Dataflow languages are extremely popular — the number of spreadsheet users alone greatly exceeds the number of programmers of any mainstream programming language — so studying them is arguably worthwhile. Semantically, so long as such languages are hosted inside safe languages, safety is not an issue. Safety is also not an issue in the case of 'pure' dataflow languages, in which the semantic model can be expressed denotationally via systems of recursive and co-recursive equations, such as in the case of Lucid Synchrone [9]. In these cases the interesting issue is efficiency, particularly of memory utilisation.

A semantically interesting question occurs in the case of languages which allow the explicit manipulation of the dataflow graph using imperative constructs. This is the case of Tensorflow [1] and of (imperative) self-adjusting computation (SAC) systems such as the Incremental library for OCaml [43]. The latter, although primarily intended for usage in high-performance computation [2], proved to be a convenient dataflow idiom for user interfaces and web programming.

Constructing a dataflow graph imperatively can be convenient, but both TensorFlow and Incremental are unsafe, in the sense that 'illegal' dataflow graphs can be constructed during the execution, which leads to unsafe behaviour. In the case of TensorFlow it is possible for the dataflow evaluation to diverge because of cyclic dependencies, whereas in the case of Incremental illegal dataflow graphs with cycles cause crashes. These problems can be avoided by a judicious language design.

In this paper we propose a programming language which is:

**imperative:** the dataflow (computation) graph is explicitly manipulated using commands;

**functional:** the language includes (call-by-value) PCF [48] as a subset;

**simply-typed:** the type system is essentially that of PCF;

**modal:** the execution of the dataflow (computation) graph is also controlled by explicit commands;

**parallel:** the dataflow graph is executed in parallel.

For such a programming language, establishing safety and memory efficiency are both challenging and non-trivial problems. Our theoretical contributions are:

**specification:** we give a precise mathematical definition of the language using an *abstract machine*, harmonising the semantics of the PCF-like language with that of dataflow;

**safety:** we show that for any well-typed program the abstract machine executes without reaching illegal states;

**deterministic:** we show that the parallel evaluation of the dataflow graph produces deterministic results;

**termination:** moreover, we show that the execution of any recursion-free program with dataflow terminates by reaching the final state of the abstract machine;

**efficiency:** we prove that no operation is 'expensive' in terms of time and memory utilisation.

**Remark.** These technical results add up to the most important contribution of the paper, which is one of *programming language design*. We demonstrated that functional and dataflow programming, resulting in the runtime creation of straight-line-code models, can be integrated seamlessly (syntactically and semantically), safely, and efficiently.

## 2 Synchronous Dataflow

Before technical details are provided let us try to understand the intended semantics of the language via some simple examples. This examples, although simple, should illustrate the versatility of the imperative style of constructing dataflow graphs. The starting language is call-by-value PCF, a typical language combining the simply-typed lambda calculus with arithmetic-logical operations and recursion. To it we add a new type of *Cells* along with facilities for manipulating them. The distinction between the type of cells and type of integers is more subtle than in languages with assignable state: the nodes in the dataflow graph have type *Cell* whereas their inputs and outputs are edges and have type *Int*. Note that in our graphs, the edges represent the dependency relations between nodes, therefore the input of a node would be represented by its outgoing edge (on top) while the output its incoming edge (at the bottom).

### 2.1 Dataflow graphs

The default mode of a program is to construct a dataflow graph, which we will describe now. The nodes of a dataflow graph are:

**Constants** have a single output edge and produce the value stored in the node.

**Arithmetic-logic operators** have two input edges and one output edge and compute the operation accordingly.

**Contraction** nodes have multiple outputs and a single input, copying their input value to the outputs. They are labelled by a $c$.

**If-then-else** have a control input and two input branches, but behave like *selectors* rather than control-flow operators.

**Cells** have a single input, which can modify the stored value, which is the output. They are labelled with their current value in curly brackets.

For example, the very simple dataflow graph (DFG) in figure 1 computes the expression *if* $y \leq x$ *then* $x$ *else* $y$, with variables (cells) initially $x = 0, \ y = 1$. The inputs of the cells are connected to the constants 2 and 3, respectively. On the first execution the DFG will produce 1 and update the cells. On any subsequent execution the DFG will produce 3. Note that we are drawing the DFG rather unusually, with the arrows indicating *data dependencies* rather than the dynamics of the data flow.

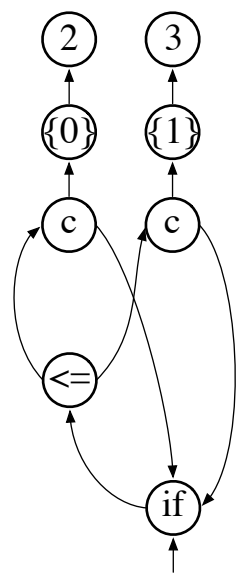

**Figure 1** If

The only non-obvious behaviour here is that of the if statement. In a conventional language it is a control-flow operator, which will evaluate first the test then only one of the branches, depending on the value of the test. In a dataflow setting both branches of the if statement compute, with the test selecting which value is propagated further down.

Crucially, what we mean by "an execution" is a propagation of values in the DFG until they either reach the output or they reach and subsequently updated every cell at most once. This is the sense in which the DFG is "synchronous", much like a digital circuit. Along computation data travels independently and possibly asynchronously, but all memory cells serve as synchronisation points. In figure 2 we can see a DFG which produces 1s and 0s, alternately, with each execution. On each execution the memory cell is updated $x' = 1 - x$.

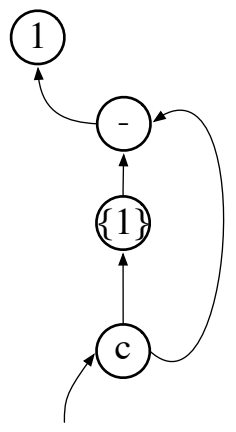

**Figure 2** Alternating

**Remark.** The drawing of the DFG is unusual in that we point the arrows in the direction that indicates data dependencies rather than information flow. The reason is that the actual execution of the DFG, discussed in section 3, is demand-driven, which each DFG node first requesting data from their dependents, which request data from their dependents, and so on. This more elaborate execution mode requires some auxiliary nodes which we have omitted in this initial, informal, presentation.

## 2.2 Graph construction and manipulation

The host language is call-by-value PCF, extended with several key new operations used to create and manipulate DFGs:

ref : *Int* → *Cell*: create and return a new cell node in the dataflow graph initialised with an integer value;

deref : *Cell* → *Int*: take a cell node in the dataflow graph as argument and produce an incoming edge as result;

root : *Cell* → *Int*: take a cell node in the dataflow graph and return its outgoing edge;

link : *Cell* → *Int* → *Unit*: take a cell node and change its outgoing edge to the edge given as an argument.

To construct programs it is convenient to also have more conventional assignment and dereferencing operations for instantly changing and reading the values of cells

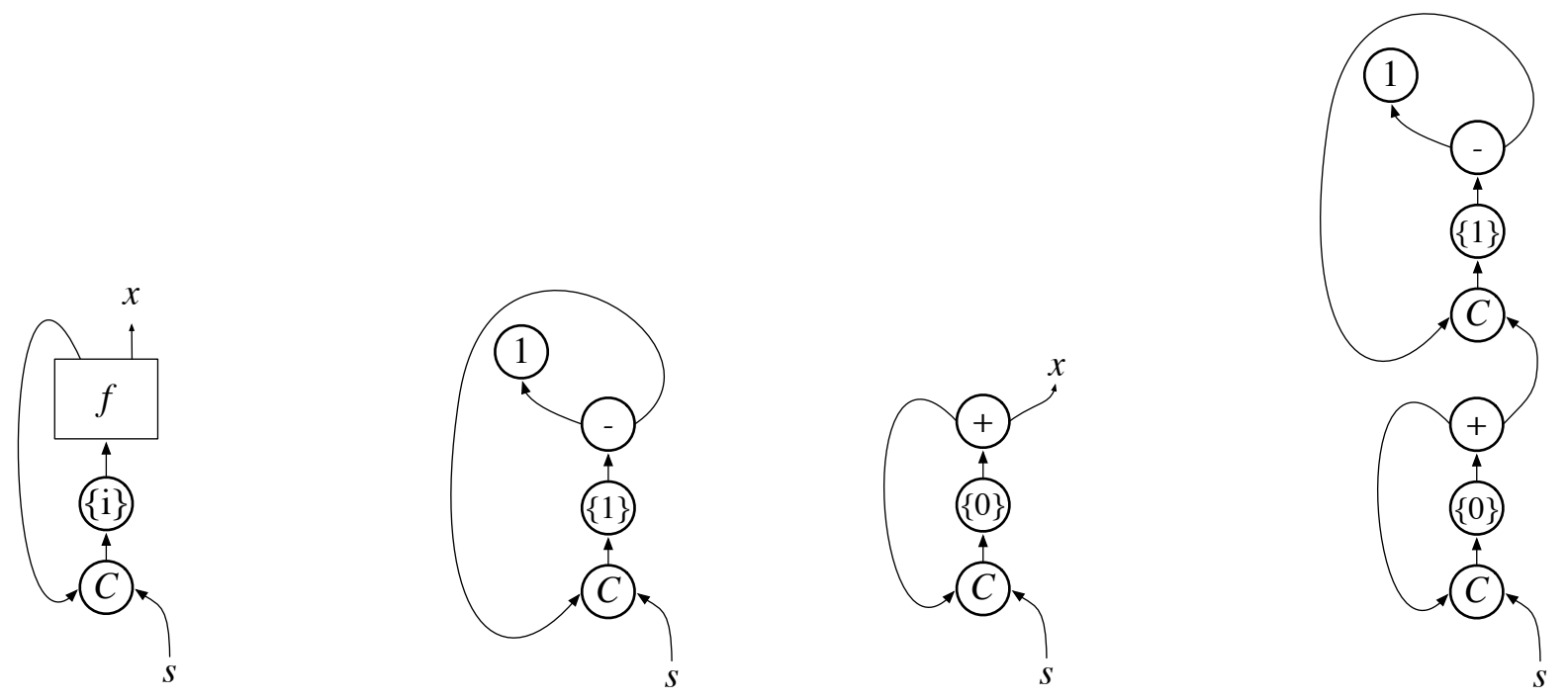

**(a)** **(b)** **(c)** **(d)**

**Figure 3** Informal examples

rather than just manipulating dependencies. In these operations the type *Integer* is interpreted in the conventional way, as a value. This dual treatment of the type of integers may seem ambiguous, but it is not, because each integer *value* can be also seen as an immutable node in the dataflow graph. Taking the value itself or the edge of the node means the same thing.

assign : $Cell \to Int \to Unit$: change the constant of a cell instantly;

peek : $Int \to Int$: extract the value of a cell instantly.

Finally the command step : *Int* activates the computation in the dataflow graph by updating the contents of each cell in parallel, and at most once, if its dependencies have changed. This allows computing in dataflow graphs with cycles, provided that each cycle contains at least one cell. In effect, the model of propagation is synchronous, with the cells serving as synchronisation points. The command will return the number of updated cells in an update cycle. Knowing the number of updated cells is a useful convenience, as it will allow repeating the propagation step until the graph 'stabilises', i.e. no more updates happen. This is normally possible if and only if the dataflow graph lacks cycles.

For the purpose of examples we consider a standard 'let' construct as syntactic sugar. Let us now consider a 'hello world' style example for dataflow computation: state machines. A generic state machine is created in our language by a function ($sm$) taking as arguments an initial value $i$, a transition function $f$, and a input $x$:

$$\mathbf{let}\, sm = \lambda i.\lambda f.\lambda x.\mathbf{let}\, s = \mathrm{ref}\, i\, \mathbf{in}\, \mathrm{link}\, s\, (f\ s\ x); \mathrm{deref}\, s$$

Because of the way $i$ is used as an argument to reference creation we can think of it as a plain integer value, whereas $f$ and $x$ should be thought of as edges in the dataflow graph. The resulting graph will have outgoing edges $x$, and incoming edge deref$s$. Besides the $i$-initialised cell-node it will also include a special 'contraction' node $C$ so that the incoming edge can be used both as output and as cyclic dependency. The function $f$ will be used, during execution, to create a desired dataflow graph. The resulting dataflow will have the general shape given in figure 3a.

Let us call the function above $sm$ and consider some simple instantiations, the examples are also available in the online visualiser[1]:

**oscillator** is a state machine which alternates between 0 and 1, defined as $\textbf{let}\, alt = sm\,1\,(\lambda s.\lambda i.1 - \text{deref}\, s)\,0$. Note that the outgoing edge is not used, and 0 is a dummy value. The created dataflow graph is represented informally in figure 3b, where the sub-graph produced when the particular transition function was supplied now has nodes for addition and the constant 1. The cell is initialised to 1.

**sum** is a state machine which accumulates the value of its inputs into the state, defined as $\textbf{let}\; sum = \lambda x.sm\; 0\; (\lambda s.\lambda i.i + \text{deref}\; s)\, x$. The dataflow graph is sketched out in figure 3c.

**composite** is a state machine which connects the incoming edge of the oscillator to the outgoing edge of the accumulator, defined as $\textbf{let}\, a = sum(\text{deref}\, alt)$. It can be seen in figure 3d.

The code above only constructs the dataflow graph. To execute it we need to execute the step command along with peek $a$ to read the output. After one *step*, for example, the first cell will change its stored value from 1 to 0, and the second one from 0 to 1. After another *step* the cells will further change to 1 and 1, respectively, and so on.

Similar programs written in Lucid Synchrone and ReactiveML are shown in appendix A.

**Remark.** Although the language is *modal*, with distinct executions for the PCF-like fragment and the DFG fragment, the syntax of the language is uniform. So the language lacks the "metaprogramming" feel of TensorFlow and other embedded DSL-like solutions. Concretely, any operator is used in the same way whether as a part of the host language or the DFG. For example, when we write $u + v$ the *semantic interpretation* is able to tease out whether this is PCF addition or DFG addition. This requires no static analysis or other syntactic solutions such as types, type classes, operator overloading and so on. This is akin to operator overloading, except that it is realised an a purely semantic way. Establishing a "semantic context", defined by its history of execution, is as far as we know a novel approach which, as we shall see, will require a novel approach to semantics.

### 2.3 Formal syntax

The calculus is an extension of the simply-typed lambda calculus with base types integers, integer cells, and unit. Let $\gamma ::= \mathit{Int} \mid \mathit{Cell} \mid \mathit{Unit}$. These are the base types. The language has simple types as defined by the grammar $\tau ::= \gamma \mid \tau \to \tau$. Terms are defined by the grammar below with \$ being arithmetic operations.

$$t ::= x \mid \lambda x.t \mid t\ t \mid n \mid op \mid \text{if } t \text{ then } t \text{ else } t \mid \text{rec } x.t$$

where $op$ ranges over arithmetic operators $+, -, \times, \div : \mathit{Int} \to \mathit{Int} \to \mathit{Int}$ and the imperative operations on dataflow graphs mentioned earlier. The typing rules of the calculus are the standard ones for the simply-typed call-by-value PCF (figure 4).

[1] https://cwtsteven.github.io/TSD-visual?ex=alt/

$$\frac{}{\Gamma, x : \tau \vdash x : \tau} \quad \frac{}{\Gamma \vdash n : Int} \quad \frac{}{\Gamma \vdash op : \tau} \quad \frac{\Gamma \vdash t' : \tau \to \tau' \quad \Gamma \vdash t : \tau}{\Gamma \vdash t'\ t : \tau'}$$

$$\frac{\Gamma, x : \tau \vdash t : \tau'}{\Gamma \vdash \lambda x.t : \tau \to \tau'} \quad \frac{\Gamma \vdash t : Int \quad \Gamma \vdash t_1 : \gamma \quad \Gamma \vdash t_1 : \gamma}{\Gamma \vdash \text{if } t \text{ then } t_1 \text{ else } t_2 : \gamma} \quad \frac{\Gamma, f : \tau \to \tau' \vdash t : \tau \to \tau'}{\Gamma \vdash \text{rec } f.t : \tau \to \tau'}$$

**Figure 4** Typing rules

# 3 Abstract machine semantics

## 3.1 The Dynamic Geometry of Interaction

We will formulate the semantics of the language using a graph-rewriting abstract machine called the *Dynamic Geometry of Interaction Machine* (DGoIM). An abstract machine semantics has the desirable feature that it can allow reasoning about the cost of computation. In fact the DGoIM machine has been shown to be optimal, in a technical sense [4], for a variety of reduction strategies for the lambda calculus [46]. Abstract machines have been generally considered unsuitable for reasoning about programming languages. However, the DGoIM has been used to define and reason about the safety of 'graph abstraction', an exotic operation inspired by TensorFlow which converts local state variables into function parameters [44]. A generalisation of the DGoIM machine has also been used to reason about observational equivalence in languages with effects [24].

The origins of the DGoIM are in token abstract machines [17] giving semantics to linear logic proofs via the 'Geometry of Interaction' [25]. Using translations of lambda calculus into its linear version these abstract machines have been also used to model PCF and related languages [39]. These models proved to be interesting as they lead to space-efficient compilation schemes which have been used, among other things, to compile functional languages into circuits [23].

The motivation for a 'dynamic' version of the machine was that the original version only gives an efficient model for call-by-name. To represent call-by-value not only soundly but also efficiently the 'token' that traverses the machine during execution also modifies it, thus preventing full re-evaluations of components which have been already evaluated.

Although the graph-rewriting-based approach of the DGoIM is new, and hence it may appear unintuitive at a first sight, it is in fact closely related to rewrites on the abstract syntax tree of the term, with a few tweaks:

**contraction:** Unlike a standard AST in which variables sit in the 'nodes' of the tree, in the DGoIM representation variables with the same name and the same scope are 'joined' using 'contraction' nodes. Moreover, when the binder of a variable occurs the edges that correspond to the variable connect back to the binder. In figure 5 you can compare the conventional AST representation of $\lambda x.x\,x$ to its DGoIM representation. The graph contains some other GoI-inspired artefacts which we discuss below.

**boxes:** In the same figure we can see, that the lambda term is 'boxed'. This can be understood as the sub-graph corresponding to the lambda term is a single entity—for purposes of copying or deleting. This is part of the machinery that allows a precise cost of computation to be associated with graph rewrite.

**linear logic:** Besides the contraction node ($C$) we can also see a !-labelled node (in grey) which is also inherited from linear logic. This node, along with ? nodes which will be seen later, serve as interface nodes to boxes, for arguments and non-local variables.

Graph rewriting in general can be expensive, in particular the identification of redexes as sub-graphs. In DGoIM this is dealt with by making the rewriting 'focussed', i.e. relative to a special node in the graph which we call 'the token'. The token will traverse the graph and trigger reductions as needed. In terms of conventional abstract machines, the movement of the token along an edge, seeking a redex, is akin to the 'narrowing' of the context in a term abstract machine, whereas the movement against the edge, after a reduction, corresponds to a 'broadening' of a term-reducing abstract machine. A visualiser is also available for demonstrating the step-by-step execution of the machine.[2]

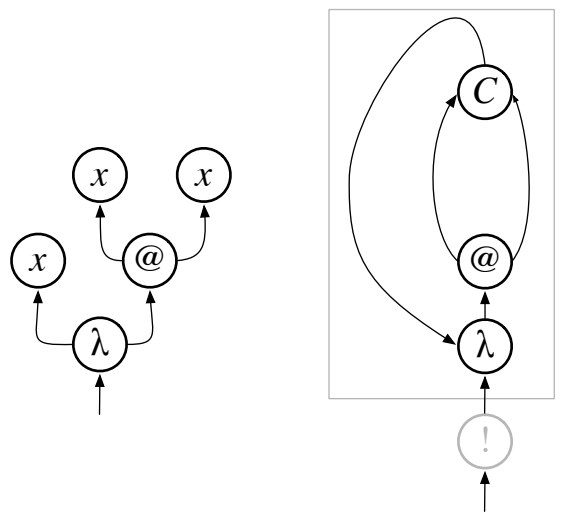

**Figure 5** AST vs DGoIM

The version of DGoIM machine presented here is essentially the original one, with the following extensions:

**Graph-construction:** Semantics for imperative operations for constructing the dataflow graph. These do not represent significant changes to the machine but are rather straightforward extensions.

**Graph-execution:** When the 'step' command is executed the program changes modes of operations, from the functional-imperative mode of execution of graph construction to the synchronous dataflow mode of execution defined by the graph. This requires a non-trivial enhancement of the DGoIM, primarily by enabling parallel execution.

### 3.2 Graph and translation

In this section we present the technical details of the DGOI machine.

A graph $G$ is a set of nodes which are partitioned into *proper* nodes $Node_G$ (or just 'nodes') and *port* nodes $Port_G$ (or just 'ports'), and a set of directed edges connecting ports and nodes. Every node is associated and connected with several ports and each edge has at least one connecting port and there are no open edges. Every port is the source of at most one edge and the target of another. We also call the source and the target of a node the incoming port (or in-port) and the outgoing port (or out-port) respectively. For example the +-node is connected to two out-ports and one in-port. If a graph contains edges between two ports they are identified by an edge homeomorphism; our graphs are quotiented by the congruence induced by edge homeomorphisms in an appropriate category of graphs [32]. We often denote a graph by $G : \Lambda \to \Gamma$ where $\Lambda$ and $\Gamma$ are the set of open incoming ports and outgoing ports respectively. They are also called the incoming and outgoing interfaces.

[2] https://cwtsteven.github.io/TSD-visual/

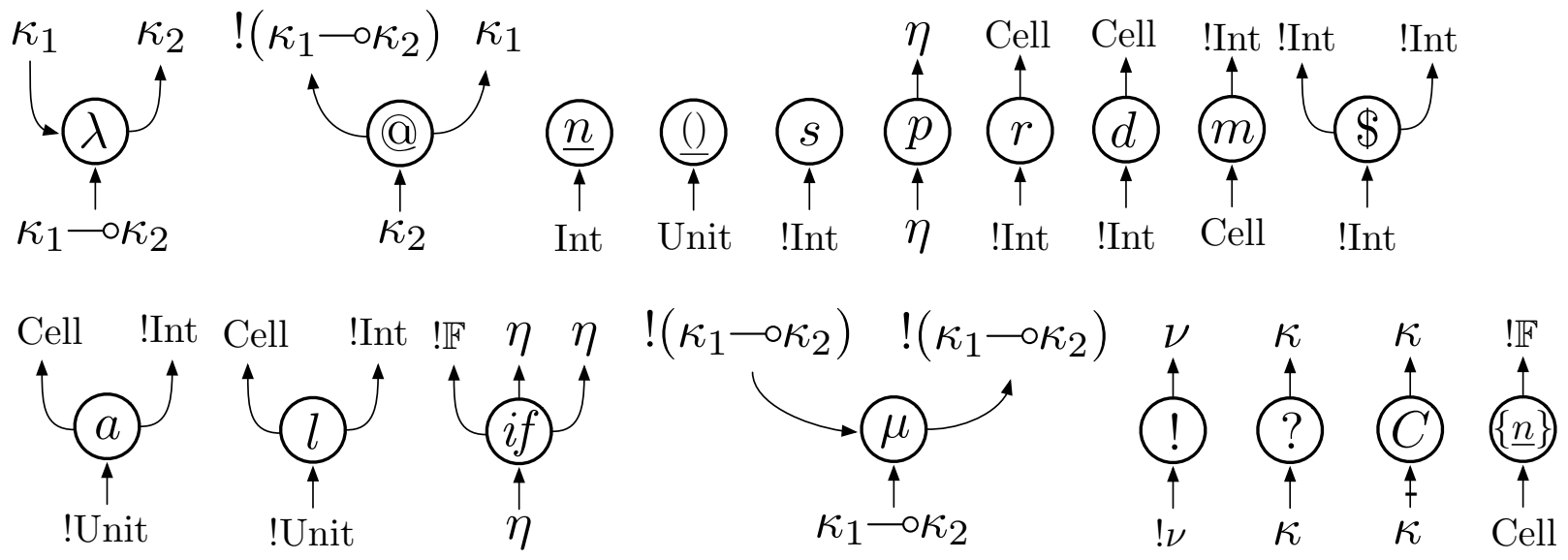

**Figure 6** Connection of edges

Ports are typed ($\theta ::= \nu \mid \kappa$), with either linear types $\nu ::= Int \mid Unit \mid \kappa \multimap \kappa$ or qualified types $\kappa ::= !\nu \mid Cell$. We also denote qualified base types as $\eta ::= !Int \mid !Unit \mid !Cell$. Adjacent ports have the same types, to be coherent with the wire homeomorphism.

We often refer to a node using their labels, for example, we say *a $\lambda$-node* for a node labelled by $\lambda$. Some labels are used to indicate nodes that interpret calculus constructs: $\lambda$-node (abstraction), @-node (application), $n$-node (constants), $s$-node (step), $p$-node (peek), $d$-node (dereference), $r$-node (root dependency), $m$-node (memory cell creation), \$-node (arithmetic operations), if-node (conditionals), $\mu$-node (recursion), $a$-node (assignment) and $l$-node (link). Some nodes do not correspond to a syntactic construct but are created at runtime to represent dataflow graphs: $\{n\}$-node (memory cell). Finally the !-node, ?-node and $C$-node (contraction) are markers used to guide sub-graph copying. These infrastructure nodes are conceptually derived from linear logic and the Geometry of Interaction (GoI) interpretation of proofs in linear logic, particularly as used in abstract token machines for GoI. The connection of edges via nodes must satisfies the typings in figure 6.

When drawing graphs, certain diagrammatic conventions are employed. Ports are not represented explicitly and their labels are only shown when they are not obvious. An edge with a bar at its tail represents zero or more ports, while a double-stroke edge represents a bunch of the same nodes and ports in parallel (figure 7).

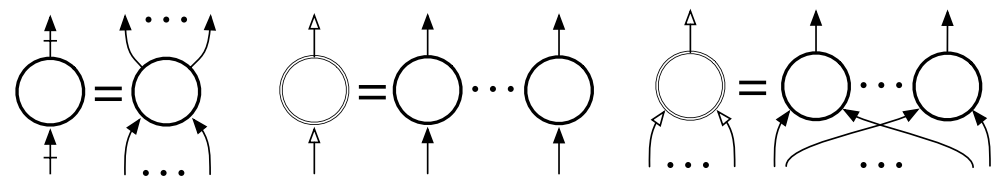

**Figure 7** Drawing conventions

We first define the translation from type judgement of a term to graphs in figure 8 with $op$ any unary and $op'$ any binary operator. The rules essentially construct the abstract syntax graph in the way discussed earlier: joining free-variable ports via contractions, looping free-variable back to the binders, inserting boxes around value-sub-graphs, adding linear logic token management nodes (!, ?). Note that all operations are eta-expanded, so, for example, instead of having just a + node in the graph we have a sub-graph corresponding to $\lambda x.\lambda y.x + y$. Finally, the $C$ node with no in-ports is the degenerate contraction (usually referred to as ‘weakening’) used to mark unused variables in a subterm.

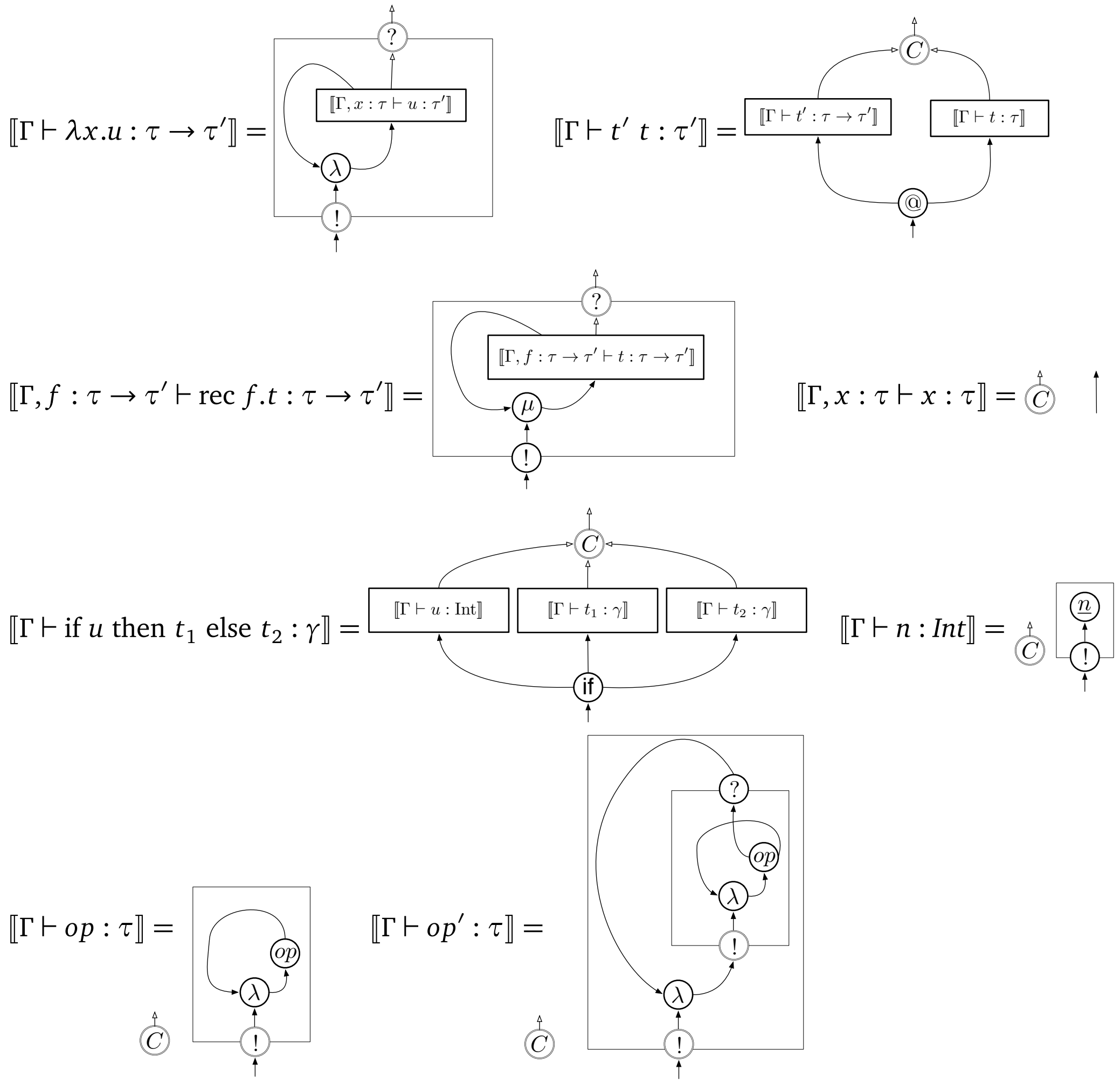

**Figure 8** Representation of terms

### 3.3 Machine states

As explained, the machine execution is given by an evaluation token traversing the graph and rewriting it. To manage the process the machine also uses data, as defined below. We call this data ‘evaluation token data’ as it can be thought of as data carried by the token.

**Definition 3.1** (Evaluation tokens)**.** For any graph $G$, the state of an evaluation token $(e,d,f,S,B)$ consists of a *port* $e \in Port_G$ indicating the token position, a *direction d*, a *rewrite flag f*, a *computation stack S* and a *box stack B* with

$$d ::= \uparrow \mid \downarrow$$
$$f ::= \square \mid @ \mid \text{if} \mid C \mid ! \mid \mu \mid m \mid p \mid l(i) \mid a(n,i) \mid r(i) \mid sp \mid s$$
$$S ::= \square \mid \star : S \mid (\lambda,-) : S \mid (n,-) : S \mid (n,g) : S \mid (n,i) : S \mid ((),-) : S \mid \text{if}_0 : S \mid \text{if}_1 : S$$
$$B ::= \square \mid i : B$$

where $i \in Port_G$.

In the definition above $e$ denotes the current *position* of the token. The *rewrite flag* determines the possible graph rewriting. The *computation stack* tracks the intermediate results of the evaluation and the *box stack* tracks duplication of values. Together the two stacks guide the transition of token. In particular the token asks the value of a graph by having $\star : S$ as its computation stack. After traversing the graph it could return either $(\lambda,-) : S$, $(n,r) : S$ or $((),-) : S$. The first part $(\lambda,n,())$ is the return value while the second part indicates whether it is a cell ($i \in Port_G$ with $i$ denoting the in-port of the cell), a graph $(g)$ or a normal value $(-)$.

**Definition 3.2** (Machine states)**.** A machine state $(G,\delta,P)$ is a triplet consisting of a *graph G*, an *eval token $\delta$* (hereinafter referred to as the main token) and a *set of eval tokens P* (hereinafter referred to as the set of prop tokens) that are used during propagation.

We define a binary relation on machine states called transitions $(G,\delta,P) \mapsto (G',\delta',P')$ which includes normal pass and rewrite transitions. We group the transitions into *pass transitions* in which only the token changes, and *rewrite transitions* in which the underlying graph may change.

### 3.4 Pass rules

The pass transitions are given in table 1. The token is at a distinguished port in the graph and the node column represents the label of one of the two nodes of the port. The second column ('e') represents whether it is an incoming $(i)$ or an outgoing $(o)$ port. We labelled the ports by its left-to-right order from zero as in figure 9. The next three columns represent the state of the direction $d$, flag $f$, and stack $S$. The next column represents the new port $e$ visited by the token, relative to the node. The final three columns represent the new values of the direction, flag, and stack.

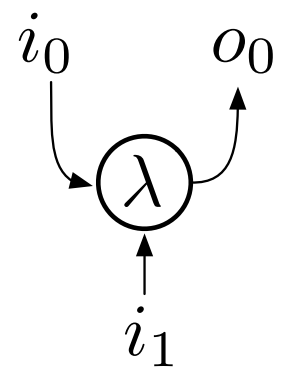

**Figure 9**

**Node $\lambda$:** If the token is at the first incoming port $i_1$ of a $\lambda$ node travelling along the direction of the edge $d = \uparrow$ the token will reverse directions $d = \downarrow$ and the top of the stack will be replaced by information that a $\lambda$ node has been encountered by placing $(\lambda,-)$ at the top. For reasons of well-formedness a $\star$ must have been at that op of $S$. This rule reflects the fact that a $\lambda$-term is a value which requires no further evaluation.

**Node** @**:** If the token reaches the sole incoming port $i_0$ of an application node it will pass through and move to its first outgoing port $o_1$ (right out-port), reflecting the right-to-left call-by-value evaluation. As it returns $d =\downarrow$ along the same port $o_1$ it is passed to the zeroth outgoing port $o_0$ (left out-port) with changed direction $d =\uparrow$ and the top of the stack become $\star$. Finally as it returns $d =\downarrow$ on the same port it is passed to the incoming port $i_0$ with $d =\downarrow$ in preparation for the rewrite corresponding to application. Node that it is expected that the last value encountered on the left side was a lambda term, since the rule expects $(\lambda, -)$ at the top. The imminence of the rewrite rule is indicated by the rewrite flag $f = @$.

**Nodes** $n, ()$**:** These nodes represent values so their behaviour is quite similar to that of the $\lambda$ node, reflecting the token and placing the value on $S$.

**Operation nodes** These nodes behave similarly to @. Incoming-port tokens are propagated to the outgoing port(s). When they come back along the outgoing port the behaviour is different according to whether they are unary or binary operators. If they are unary the token is passed to the incoming port and the rewrite flag $f$ is used to indicate an imminent rewrite. If they are binary then the second outgoing port is visited first.

**Node** $\mathit{if}$**:** This node behaves like a ternary operator, except that after returning from the first outgoing edge the top of the stack is inspected.

**Node** $\mu$**:** The recursion node immediately prepares the machine for a rewrite by changing the flag to $f = \mu$.

**Node** !**:** This node, also found in conventional GOI abstract machines, indicates that a box is about to be processed. The rewrite rule about to be triggered is indicated by the flag becoming $f = !$.

**Node** $\{\underline{n}\}$**:** The cell storing value $n$ is dealt with almost as a standard integer value. The only difference is that the port $i_o$ is also placed on top of $S$ along with the value $n$, to be possibly used by datflow graph operations.

**Node** $C$**:** Contraction nodes use the other stack, the box-stack, to remember along which of the several incoming ports a token arrived so that it can return to the same port. Moreover, the incoming port prepares a rewrite $f = C$, essentially the copying of the shared term (represented as a subgraph).

Note that certain nodes, such as ?, have no associated rules because the token is never supposed to reach them.

### 3.5 Rewrite rules

The pass rules only serve to set the scene for the rewrite transitions, which are given in figure 10.

**The *Beta* rule:** The essential computational rule of the lambda calculus is the removal of the abstraction-application pair. The second outgoing port of the application node @ is connected to the argument, which is now re-wired to the first argument of the subgraph $G$ representing the function body. This is a 'small' beta law since it involves no copying of the argument. To anticipate a little, the copying will be performed by the contraction nodes.

**Table 1** Pass transitions

| node | e | d | f | S | $\mapsto$ | e | d | f | S |
|---|---|---|---|---|---|---|---|---|---|
| $\lambda$ | $i_1$ | $\uparrow$ | $\square$ | $\star$:S | | $i_1$ | $\downarrow$ | $\square$ | $(\lambda, -)$:S |
| @ | $i_0$ | $\uparrow$ | $\square$ | S | | $o_1$ | $\uparrow$ | $\square$ | S |
| @ | $o_1$ | $\downarrow$ | $\square$ | X:S | | $o_0$ | $\uparrow$ | $\square$ | $\star$:S |
| @ | $o_0$ | $\downarrow$ | $\square$ | $(\lambda, -)$:S | | $i_0$ | $\uparrow$ | @ | $\star$:S |
| $n$ | $i_0$ | $\uparrow$ | $\square$ | $\star$:S | | $i_0$ | $\downarrow$ | $\square$ | $(n, -)$:S |
| () | $i_0$ | $\uparrow$ | $\square$ | $\star$:S | | $i_0$ | $\downarrow$ | $\square$ | $((), -)$:S |
| For $\omega \in p, d, r, m$ (unary operations): | | | | | | | | | |
| $\omega$ | $i_0$ | $\uparrow$ | $\square$ | S | | $o_0$ | $\uparrow$ | $\square$ | S |
| p | $o_0$ | $\downarrow$ | $\square$ | $(n, x)$:S | | $i_0$ | $\downarrow$ | $p$ | $(n, -)$:S |
| r | $o_0$ | $\downarrow$ | $\square$ | $(n, i)$:S | | $i_0$ | $\downarrow$ | $r(i)$ | $(n, -)$:S |
| m | $o_0$ | $\downarrow$ | $\square$ | $(n, x)$:S | | $i_0$ | $\downarrow$ | $m$ | $(n, -)$:S |
| For $\# \in \$ \cup a, l$ (binary operations): | | | | | | | | | |
| # | $i_0$ | $\uparrow$ | $\square$ | S | | $o_1$ | $\uparrow$ | $\square$ | S |
| # | $o_1$ | $\downarrow$ | $\square$ | S | | $o_0$ | $\uparrow$ | $\square$ | $\star$:S |
| $ | $o_0$ | $\downarrow$ | $\square$ | $(m, -) : (n, -)$:S | | $i_0$ | $\downarrow$ | $ | $(\$\ m\ n, -)$:S |
| a | $o_0$ | $\downarrow$ | $\square$ | $(m, i) : (n, x)$:S | | $i_0$ | $\downarrow$ | $a(n, i)$ | $((), -)$:S |
| l | $o_0$ | $\downarrow$ | $\square$ | $(m, i) : (n, x)$:S | | $i_0$ | $\downarrow$ | $l(i)$ | $((), -)$:S |
| if | $i_0$ | $\uparrow$ | $\square$ | S | | $o_0$ | $\uparrow$ | $\square$ | S |
| if | $o_0$ | $\downarrow$ | $\square$ | $(0, -)$:S | | $o_1$ | $\uparrow$ | $if$ | $\star$:S |
| if | $o_0$ | $\downarrow$ | $\square$ | $(n \neq 0, -)$:S | | $o_2$ | $\uparrow$ | $if$ | $\star$:S |
| if | $o_1$ | $\downarrow$ | $\square$ | S | | $o_2$ | $\uparrow$ | $\square$ | $\star$:S |
| $\mu$ | $i_1$ | $\uparrow$ | $\square$ | S | | $i_1$ | $\uparrow$ | $\mu$ | S |
| ! | $i_0$ | $\uparrow$ | $\square$ | S | | $o_0$ | $\uparrow$ | ! | S |
| ! | $o_0$ | $\downarrow$ | $\square$ | S | | $i_0$ | $\downarrow$ | $\square$ | S |
| $\{n\}$ | $i_0$ | $\uparrow$ | $\square$ | $\star$:S | | $i_0$ | $\downarrow$ | $\square$ | $(n, i_0)$:S |
| where $x \in \{-, g\}$ | | | | | | | | | |

| node | e | d | f | S | B | $\mapsto$ | e | d | f | S | B |
|---|---|---|---|---|---|---|---|---|---|---|---|
| C | $i_k$ | $\uparrow$ | $\square$ | S | B | | $o_0$ | $\uparrow$ | $C$ | S | $i_k$:B |
| C | $o_0$ | $\downarrow$ | $\square$ | S | $i_k$:B | | $i_k$ | $\downarrow$ | $\square$ | S | B |

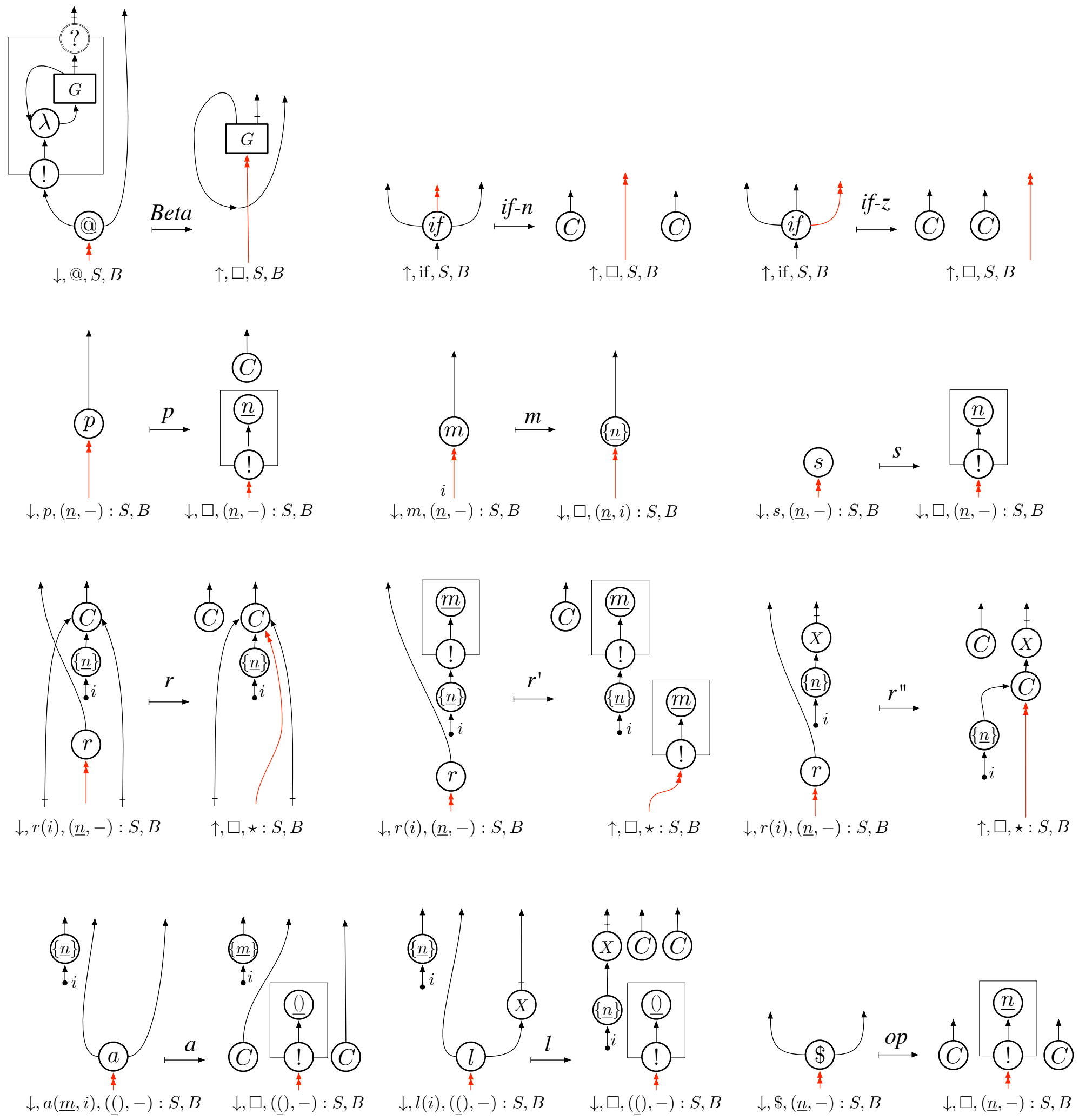

**Figure 10** Computation rewrites

**The $s$-rule:** represents the 'step' command which evaluates the dataflow graph, which will be discussed later. This rewrite, with an token exiting the incoming port $d = \downarrow$ is performed after this evaluation. The dataflow evaluation leaves a value $n$ on $S$ indicating that $n$ nodes were updated, so the $s$-node is replaced by a constant $n$ value box.

**The $p$-rule:** represents the 'peek' command which reads the value off the dataflow graph, as pointed at by the $p$ node, and creates an constant $n$ valued box from it. Note that the outgoing port of the $p$ node is replaced by a weakening $C$, indicating that port became unreachable.

**The $m$-rule:** represents cell creation, and is executed by replacing the $m$ node by a $n$-initialised cell node. The value $n$ is retrieved off the stack $n$, where it is augmented

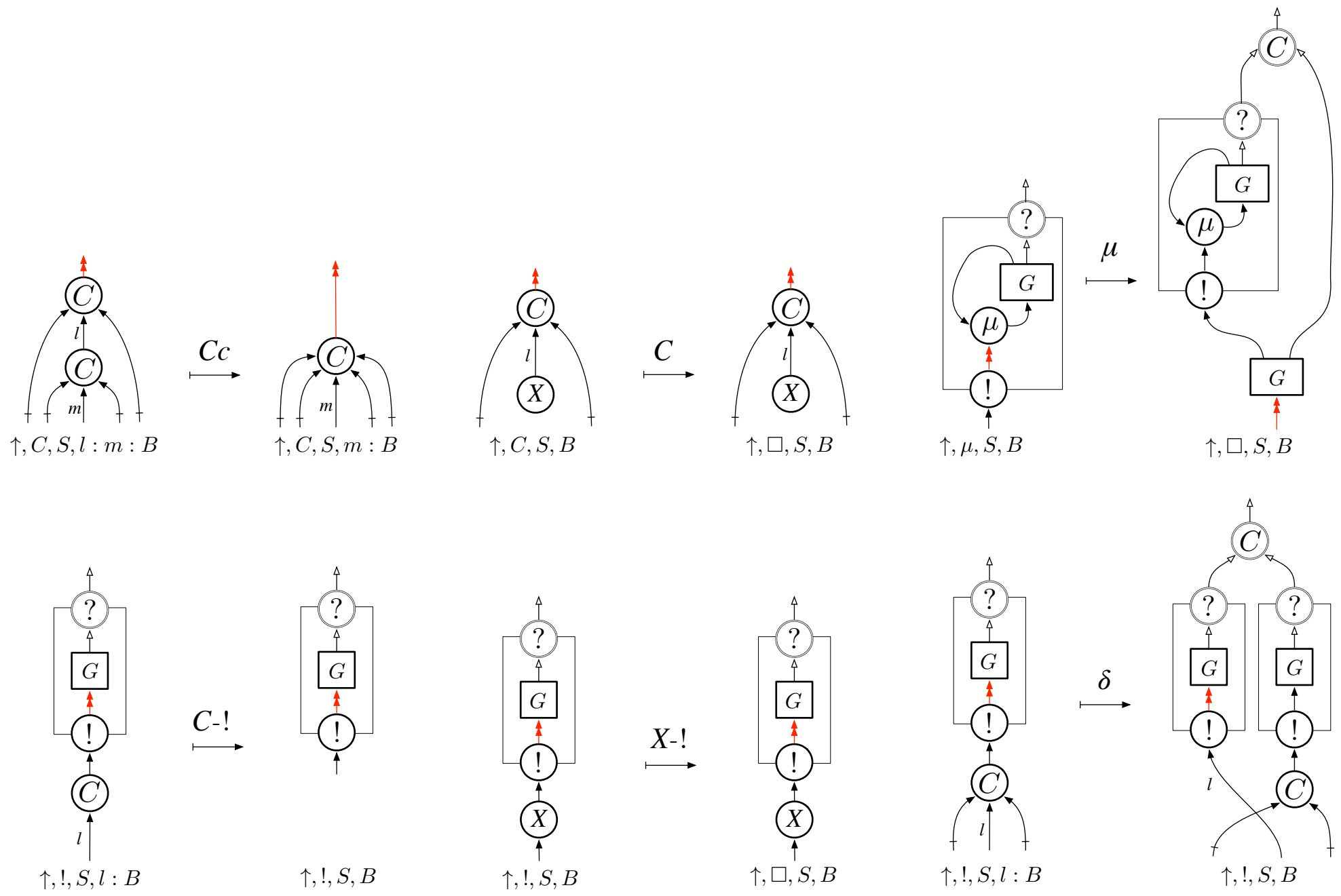

**Figure 11** Copying rewrites

with the port $i$ so that it can be used by other dataflow graph manipulating operations.

**The $r$-rules:** compute the dependency ('root') of a graph. There are three versions of the rule $(r, r', r'')$ depending on whether the requested dependency is incoming to a contraction, constant, or some other node. In all cases the token 'jumps' to the incoming port of the cell, labelled by $i$. Note that this rule is 'non-local', involving edge $i$ located in some arbitrary point in the graph. The node which is at the nock of the arrow is irrelevant and it is not shown.

**The $op$-rules:** compute operators by replacing the operation node with the resulting value as returned by the token and left on $S$. This is very similar to how stack machines generally evaluate operators. The outgoing ports of the operator are detached from the graph via weakening.

**The $a$-rule:** implements assignment. A port $i$ is left at the top of $S$ along with a computed integer $m$ (in the form $(m, i)$). The result of the assignment operation is unit () but the value of the cell with incoming port $x$ is updated to $m$.

**The $l$-rule:** implements linking. At the top of $S$ there is a port $x$ (tagged as $l(i)$). The operation returns (), like assignment, but the outgoing port of the cell pointed at by port $i$ is reassigned to the second argument $i$ of $a$.

**The $if$-rule:** depending on the previous pass rule, if the token is sent to the first branch the second branch is disconnected by linking the port to a weakening, and vice-versa. In both cases the node *if* is removed.

Finally, copying rules are given in figure 11.

**The *Cc* rule:** combines two consecutive $m$-ary and $n$-ary contraction nodes in a single $m+n$-ary contraction node.

**The *C* rule:** (for a node $X \neq C$) resets the $f$ flag. This is not a 'pass' rule, even though the underlying graph does not change, because the token stays put. This happens with some other rules for contraction.

**The *C-!* rule:** removes redundant 'unary' contractions.

**The *X-!* rule:** (for a node $x \neq C$) resets the $f$ flag.

**The $\delta$ rule:** is the the main rule of the contraction node, and it leads to the copying of the box just above it. Note that this is the main purpose of boxes, to delimit copyable sub-graphs for copying by the contraction node.

**The $\mu$-rule:** is not a contraction rule, but it is very similar. It represents the 'unfolding' of a recursive call by copying the body $G$.

### 3.6 Dataflow computation

In contrast to the graphs representing programs, graphs representing dataflow are much simpler and includes only nodes for operators ($), if statements, and dereferencing (d). The pass and rewrite rules are given in table 2. To indicate the datflow-mode for operators the flag $g$ is placed at the top of the stack.

**The $d$ pass rule:** represents the dereferencing operator, which records on the stack $S$ the value $n$ stored at a cell $i$.

**The \$-operator pass rule:** picks up the values of the branches $m, n$ from the top of the stack and replace them with the value of the operator as applied to them $\$\ m\ n$. The tags $x, y \in \{-, g\}$ but at least one should be $g$. This is because we want to mix constant nodes into the dataflow graph, and they result in a $-$ tag, whereas cells result in a $g$ tag. If both of them are $-$ it means that there is a sub-graph made just of constants, which should have been evaluated during graph construction.

**The *if* pass rule:** implements the behaviour of branching in the datflow graph. Unlike an if statement in the graph-construction mode, which chooses one branch, in dataflow mode both branches are evaluated and only one value is selected, as seen in the final two rules, where $X, Y$ represent any element of the stack.

Additionally, some of the pass rules from table 1 apply to dataflow computation as well, unchanged (namely constants $n$, cells $\{n\}$, operators $\omega$, if $\uparrow$ and !).

The if-rule for dataflow is inefficient, but it is required for soundness because of conditional graph constructions. Consider the code:

$$\textbf{let } x = \text{ref } 1 \textbf{ in } \text{ref}(\text{if } x < 2 \text{ then } f\ 3 \text{ else } f\ 4)$$

If only one side is evaluated then when $x$ is possibly changed to 2 then $f\ 4$ would need to be evaluated. So instead of rewriting the if node, because it is part of a dataflow graph, it is left in the dataflow graph under the proviso that we always need to evaluate both sides. An optimised version of the rule which only evaluates one branch could be written, but it is more complicated. Moreover, the inefficiencies introduced here do not change the asymptotic complexity of graph evaluation, as will be seen shortly.

**Table 2** Dataflow rules

| node | e | d | f | S | $\mapsto$ | e | d | f | S |
|---|---|---|---|---|---|---|---|---|---|
| d | $o_0$ | $\downarrow$ | $\square$ | $(n,i)$:S | | $i_0$ | $\downarrow$ | $\square$ | $(n,g)$:S |
| \$ | $o_0$ | $\downarrow$ | \$ | $(m,x):(n,y)$:S | | $i_0$ | $\downarrow$ | \$ | $(\$\ m\ n,g)$:S |
| if | $o_0$ | $\downarrow$ | $\square$ | $(0,g)$:S | | $o_1$ | $\uparrow$ | $\square$ | $\star$ : if$_0$:S |
| if | $o_0$ | $\downarrow$ | $\square$ | $(n \neq 0,g)$:S | | $o_1$ | $\uparrow$ | $\square$ | $\star$ : if$_1$:S |
| if | $o_2$ | $\downarrow$ | $\square$ | $X : Y$ : if$_0$:S | | $i_0$ | $\downarrow$ | $\square$ | $X$:S |
| if | $o_2$ | $\downarrow$ | $\square$ | $X : Y$ : if$_1$:S | | $i_0$ | $\downarrow$ | $\square$ | $Y$:S |
| \$ | $i_0$ | $\downarrow$ | \$ | $(n,g)$:S | | $i_0$ | $\downarrow$ | $\square$ | $(n,g)$:S |

For evaluating dataflow graphs we use a different set of tokens called *prop-tokens*. Each such evaluation starts with an initial set of initial prop tokens and concludes with a set of *final prop tokens*.

**Definition 3.3** (Initial prop-tokens). We define the set of initial prop tokens $Init^P$ as $\{(e_0,\uparrow,\square,\star:\square,\square),...(e_n,\uparrow,\square,\star:\square,\square)\}$, where $\{e_0...e_n\}$ is the set of outputs of all $\{n\}$-nodes in the graph.

**Definition 3.4** (Final prop-tokens). We define the set of final prop tokens $Final^P(\vec{X})$ as $\{(e_0,\downarrow,\square,X_0:\square,\square)...(e_n,\downarrow,\square,X_n:\square,\square)\}$, where $\{e_0...e_n\}$ is the set of outputs of all $\{n\}$-nodes in the graph, $\vec{X} = \{X_0,...,X_n\}$ and $X_0,...,X_n$ are elements of the computation stack.

When the token hits a **step** node, the machine switches from evaluation mode to propagation mode.

**Definition 3.5** (Mode switching). We define the following special transitions:

1. $(G,(e,\uparrow,\square,\star : S,B),\emptyset) \mapsto (G,(e,\uparrow,sp,\star : S,B),Init^P)$ triggering the propagation when $e$ is the incoming port of an $s$-node , and
2. $(G,(e,\uparrow,sp,\star:S,B),Final^P(\vec{X})) \mapsto^s (G',(e,\downarrow,s,(b,-):S,B),\emptyset)$ terminating the propagation as follows: for any $\{n\}$-node in $G$, let the return value of the eval token sitting on its output be $(n',x)$, then if $n \neq n'$, the $\{n\}$-node is updated to $\{n'\}$ in $G'$. The value of $b$ is $1$ if there are updates and 0 otherwise where $1,0 \in Int$.

**Definition 3.6** (Propagations). We define a *propagation* as any transition from $(G,\delta,P)$ picking any non-final prop token to proceed using any possible transitions.

**Definition 3.7** (Initial and final states). For any graph $G : \Lambda \to \Gamma$, any $e \in dom(\Lambda)$ and any element $X$ in the computation stack, we define the *initial* states and *final* states as $Init(G,e) = (G,e,\uparrow,\square,\star:\square,\square,\emptyset)$ and $Final(G,e,X) = (G,e,\downarrow,\square,X:\square,\square,\emptyset)$ respectively.

The DGOI machine executes terms by encoding them into computation graphs according to the definitions in figure 8, where the translation of a well-typed term, is denoted by $[\![\Gamma \vdash t : \tau]\!]$. An *execution* is any sequence of transitions starting from the initial state. An execution is said to *terminate* if the execution ends with the final state, i.e. $Init(G,r) \mapsto^* Final(G',r,X)$ for some $G',X$. An execution is said to be *safe* if it terminates or at any state there is an applicable rule. They are formally defined in appendix D.10.

**Remark.** Much like the language itself, the graph-rewriting semantics is bi-modal, with pass and rewrite rules. The pass rules, even though they do not change the underlying graph, can still change the token. By inspecting the value of the token the semantics realises a kind of novel "semantic contextualisation" which allows certain rules to be applied in a different way depending on the execution history of the program. For example, operators can have either a pass or a rewriting semantics depending on the data in the token, which records whether the operator must be reduced or kept into a DFG graph. This is the "secret sauce" which allows operators to behave in a contextual way without resorting to any syntactic trickery.

### 3.7 Overview of type soundness proofs

In the Appendix we prove a basic theorem of the calculus, namely the Type Soundness Theorem which states that a well-typed program will not crash.

**Theorem 3.1** (Type soundness)**.** *Let* $G : \{r : [\![\tau]\!]\} \to \emptyset = [\![\vdash t : \tau]\!]$ *for some closed well-typed term* $\vdash t : \tau$*. (1) If* $t$ *is recursion-free then any execution from* $Init(G,r)$ *terminates. (2) Otherwise, the execution of* $t$ *is at least safe.*

In order to prove (1) termination, we first identify a special form of graph, dataflow graph. Any execution that runs on them will always terminate and the graph will remain unchanged since no rewrite ever happens. Throughout the execution, any cell will always be connected to a dataflow graph because of the nature of call-by-value. Therefore step-propagation always terminates and is confluent. As a result, all the non-deterministic propagation sequences can be seen as one single transition and thus making the machine deterministic. After that we can prove termination by using induction on type derivation. As for (2) safety, the proof is simply a case analysis on possible states. All the proof details can be found in appendices D.1 to D.10.

Given the formulation of the semantics as an abstract machine we can immediately conclude that:

**Theorem 3.2** (Efficiency)**.** *The following operations can be executed in linear time on the depth of the dataflow graph: cell creation, dereferencing, peeking, linking, assignment, root. The* step *operation can be executed in linear time on the depth of the dataflow graph and on the number of cells.*

### 3.8 A non-trivial example

We conclude with a non-trivial example, computing a stream of prime numbers using the sieve of Eratosthenes. The code is in listing 1 and can be executed in the visualiser.[3] It is interesting to contrast our code with the same program implemented

[3] https://cwtsteven.github.io/TSD-visual?ex=primes

**Listing 1** A stream of prime numbers

```
let fromn = λn.let s = ref n in link s (s + 1); deref s in
let filter = λinp. λn.i == n || ((i % n) <> 0) in
let inp = fromn 2 in
let sieve = ref (filter inp 2) in
let next = λ_. step; link sieve (root sieve && (filter inp (peek inp))) in
let delay = ref inp in
let primes = if deref sieve then delay else 0 in
next 0; peek primes; // return 2
next 0; peek primes; // return 3
next 0; peek primes // return 0
```

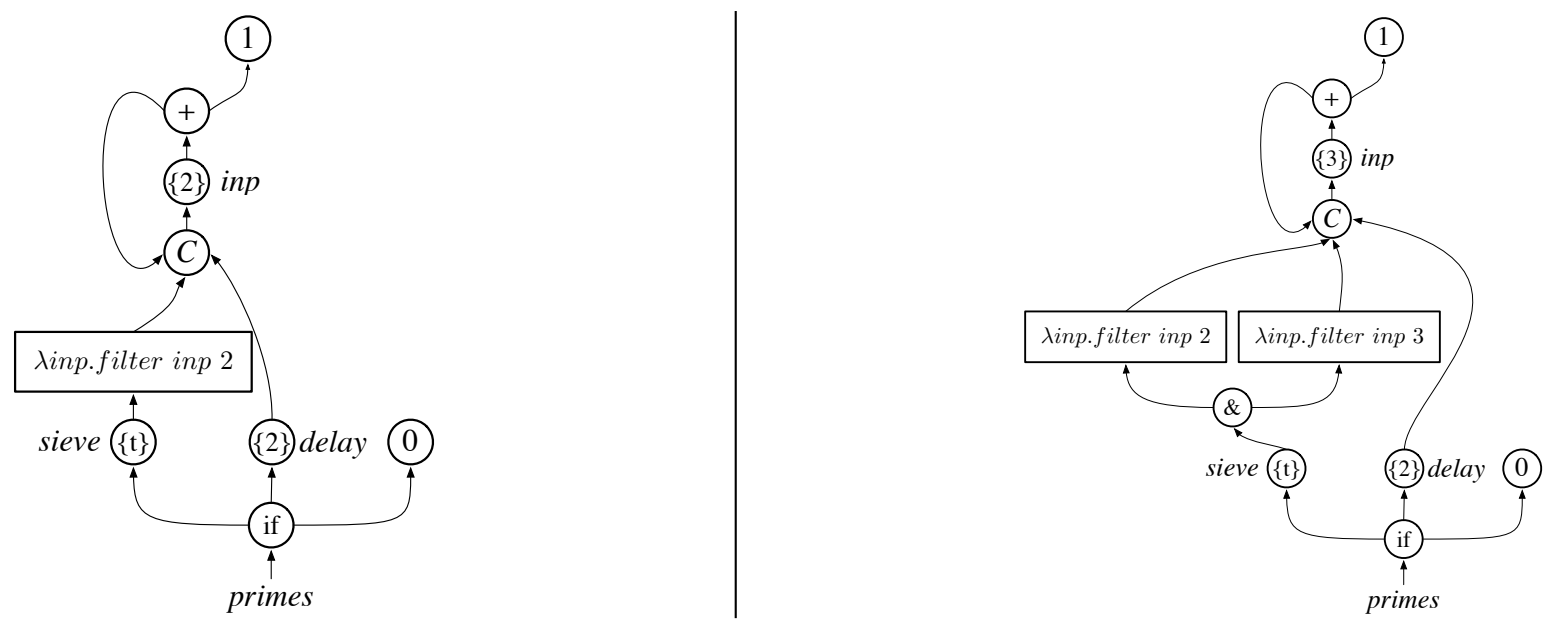

**Figure 12** before and after next()

in a similar language, Lucid Synchrone[4] (appendix B). The main distinction is that we use imperative commands to set up the dataflow graph, whereas in Lucid Synchrone co-recursive stream equations are used to the same purpose. It is a matter of preference which one of the two versions are seen as more accessible.

In listing 1, the function *fromn* takes an integer *n* and creates a stream of integers from *n*. The function *filter* takes an input stream *inp* and an integer *n* and creates a stream of booleans that could filter out the multiples of *n* (except for *n* itself) and anything smaller than *n* from *inp*. This filter function is meant to create the different layers of sieve that are used in generating prime numbers.

To implement the sieve of Eratosthenes, we first start with an input stream using *fromn* 2 (named *inp*). A cell is then initialised with the first layer *filter inp 2*. The sieve needs to expand by adding new layers of sieve in every step, without rewriting the dataflow graph. The solution is to use the function *next* which unfolds a layer after each *step* as shown in figure 12. Finally since *sieve* is a cell that depends on *inp* meaning that it is always one step behind *inp* and thus *primes = if sieve then inp else 0* would not gives the correct stream. Instead we need to create a delayed stream from *inp* (*delay = ref inp*) in order to synchronise the input of *primes*, i.e. *primes = if sieve then delay else 0*. Table 3 shows the first few elements of each stream.

[4] https://www.di.ens.fr/~pouzet/lucid-synchrone/manual_html/manual015.html

**Table 3** Streams of the sieve program

| step | 0 | 1 | 2 | 3 | 4 | 5 | 6 | 7 | ⋯ |
|---|---|---|---|---|---|---|---|---|---|
| inp | 2 | 3 | 4 | 5 | 6 | 7 | 8 | 9 | ⋯ |
| sieve | t | t | t | f | t | f | t | f | ⋯ |
| delay | 2 | 2 | 3 | 4 | 5 | 6 | 7 | 8 | ⋯ |
| primes | 2 | 2 | 3 | 0 | 5 | 0 | 7 | 0 | ⋯ |

## 4 Experimental implementation

We have proved that TSD is efficient "in principle" by examining its abstract machine execution. But we also want to gather some practical evidence that the mechanism we propose is not hopelessly inefficient. For that, we implement the calculus as a library module for OCaml (TSD), together with a PPX language extension.[5] The library serves as an embedded DSL, which is blended into the host language more seamlessly via the language extension. Note that the semantics defined earlier is used as a specification, and is not realised faithfully by the implementation. As such, the OCaml implementation does not benefit fully from the potential to seamlessly integrate the DFG and functional syntax and semantics. A full implementation based directly on the abstract machine is a more substantial endeavour which we leave as further work. Details of the implementation are in appendix C, where we also give comparisons between TSD and related programming languages, both in terms of programming style and in terms of benchmarks (appendix C.1).

## 5 Related work

The broadest context of our work is dataflow programming, which has been extensively studied ([31] is a relatively recent, exhaustive survey). Besides the numerous FRP frameworks, some of the ideas we present here are particularly prominent in functional hardware-description languages which allow (meta) programming synchronous dataflow, including in a transparent style if so desired [5]. Advances in dataflow programming materialised into the success of notable languages such as Lucid [53] or Lustre [28], but the influence of the dataflow style of programming was far broader, as seen for example in the popular machine-learning library TensorFlow, which is presented commonly as a embedded domain specific dataflow language (EDSL) in Python [1].

**Functional reactive programming** We are particularly concerned with dataflow programming in a functional style, the so-called functional reactive programming [18, 30, 47, 54] which proved to be especially well suited for Haskell because of its lazy style of computation, its expressive type system, and syntactic features which allows the convenient design and implementation of EDSLs ([8] is a recent textbook).

[5] https://github.com/cwtsteven/TSD

*FrTime* [15] is a functional reactive programming language embedded into Scheme. It also provides transparent graph creation by inserting code for operators such as {+, -} to check in runtime if their arguments are signals or ordinary Scheme values. This makes use of the fact that Scheme is dynamically typed. The graph creation mechanism in TSD is very much similar but the choice of creating a graph or to create a normal constant is captured in the rewriting rules of the node instead. It can be seen as one of the implementation techniques of transparent graph creation. There is also a difference to the model of change propagation between them. FrTime embraces a push-driven implementation where signal changes are pushed to the dataflow network immediately. On the contrary, TSD can be seen as a mixed of push and pull implementation. For immediate dependency (without cells) they are pull-driven and their values are recomputed on demand. Whereas for cells, they will be updated via *step* which is push-driven. By having explicit *step* command and pull-driven immediate dependencies, we avoid the glitch problem that arises in all push-driven implementations. *Flapjax* [42] is another FRP language based on FrTime but it is embedded into JavaScript. *AmbientTalk/R* [11] is another push-based reactive programming framework similar to FrTime that is built on top of AmbientTalk. Like FrTime, expressions in AmbientTalk/R are implicitly lifted such that graphs are created transparently. AmbientTalk/R also supports distributed reactive programming where reactive values can be distributed in various hosts. We are also interested in designing a distributed model for TSD where cells can be allocated in different machines but this is left for further investigation.

*REScala* [52] is an EDSL built on top of Scala that bridges between object-oriented and function style in reactive programming. It provides two abstractions for reactive programming, *signals* and *events*. The input of dataflow graph in REScala is called a `Var` which can be `set` or `transform` to other values. Signals are created using `Vars` and other signals. Whereas in TSD, cells are not just the input of a graph but it also represents a state in the graph. The dependency of cells can also be reconfigured using `link` which allows the graph to be cyclic. On the other hand, cyclic dependencies will lead to runtime error in REScala. Moreover, change propagation in REScala is also push-based.

A somewhat parallel but relevant line of research is the use of temporal type systems to restrict reactive computation to (space) efficient programs [33, 34, 35]. In this approach the type of streams remains exposed, so the programmer can write more expressive, lower-level (in the sense of controlling events and streams) programs, but such programs must conform to a typing discipline which guarantees certain space-utilisation constraints. In contrast, our approach is *higher-level*— less expressive but more convenient for the unsophisticated programmer. In TSD the complexity is pushed down to the level of the abstract machine, whereas in *loc. cit.* a complex temporal typing discipline is imposed on a conventional operational semantics.

**Self-adjusting computation** Self-adjusting computation has been introduced by Acar [3] and studied extensively in recent years. The use of SAC as a transparent dataflow idiom, which features quite prominently in our motivating introduction, has not been studied formally, although it has been mentioned [29] and it has been extensively discussed in informal fora.[6] In terms of implementation, SAC is a good match for *incremental computation* [49], which is a sophisticated memoisation infrastructure which allows the re-evaluation of expressions that need to be re-evaluated, thus saving computation (see also [50] for a survey). As most optimisations, incremental computation relies on a trade-off between saving the cost of unnecessary re-evaluations on the one hand versus maintaining the memoisation infrastructure, which is not free [10]. We are providing two implementations, incremental and non-incremental, and indeed the performance gains are quite dependent on the particular application.

**Synchronous dataflow languages** Also related is the significant body of work on synchronous dataflow programming languages and their semantics (e.g. Esterel [7]) and in particular functional languages (e.g. Lucid Synchrone [13], a functional version of Lustre [28]). These are the most similar languages to TSD, with some significant distinctions: TSD constructs the dataflow model using imperative features, and it does not aim at real-time execution, so it does not employ a clock calculus. Also, TSD dataflow graphs are productive by construction, since cyclic links can only be constructed via cells. So productivity checks are not necessary.

RML [41] is a low level synchronous extension to OCaml, consisting of a set of primitives for sending and handling messages. We say 'low level' because it is very easy to write unsafe programs in RML, e.g. two processes waiting for each other to emit a signal. Reactive ML has in common with the TSD language the fact that they build on the ML language, as a stand-alone compiler in their case versus a PPX extension in ours. Our work is conceptually related to such languages in the way that SAC is related to (asynchronous) FRP.

**Geometry of interaction** The DGOIM is a graph-rewriting system based on the Geometry of Interaction (GOI) [25], an interpretation of proofs in linear logic as *proof nets* [26]. The GOI was formulated as a compositional operational semantics of communicating transducers and used to interpret higher-order functional programming languages [39]. The GOI approach turned out to be particularly useful for compiling conventional languages to *unconventional* architectures (such as reconfigurable circuits [23], distributed [21], or quantum architectures [36]). It has also been used to give strikingly space-efficient compilation schemes [38] and improved time-efficiency by incorporating graph rewriting in addition to token-passing [19, 20]. The extension of the GOI machine with simple rewriting rules, the so-called Dynamic GOI machine has allowed the definition of an efficient (space and time) abstract machine for CBN, CBV and lazy (a la Haskell) evaluations [45]. This flexibility and uniformity in representing several evaluation strategies in the same machine allows a natural specification for languages that require it, such as TSD. Moreover, the computational intuitions of the approach combine seamlessly dataflow and higher-order features, making it suitable for the study of languages that require it, such as TensorFlow [14, 44].

[6] See e. g. https://blog.janestreet.com/introducing-incremental/

## 6 Conclusion

In this paper we have introduced a new programming idiom, *transparent synchronous dataflow programming* (TSD). This programming idiom is natural, allowing a high-level of abstraction to be applied to useful synchronous algorithms.

We define TSD using the relatively new diagrammatic operational style of token-guided graph rewriting. Although inspired by the Geometry of Interaction this *dynamic* mechanism is in many a significant departure from the strictures of GOI. It is arguably intuitive, as the graphical formalism is a natural setting for the study of dataflow, which it neatly extends with higher-order features. Moreover, it has certain technical advantages, such as the fine-grained management of sharing or the decomposition of *big* syntactic reduction rules into atomic graph manipulations.

The calculus TSD restricts cells to ground type. In order for multi-token (parallel) propagation to work dataflow graphs must be also *pure* in the sense that no rewrites are possible. Our solution is to prevent the presence of lambda-nodes in computation graphs. This is done for the sake of simplicity. We already know that DGOIM can be used to give CBN evaluation on a fixed graph including lambda-nodes, so an extension to higher-order TSD is likely to be possible. On the other hand, extensions with effects are more problematic as the extension of DGOIM with effects, in general, is a matter of active research.

Our PPX language extension to OCaml implementing the calculus does not take advantage of parallel evaluation of propagation or optimisation beyond basic incrementalisation. It is mainly a proof-of-concept implementation that shows that the ideas are implementable (even naively). But efficient implementations are high on our agenda.

**Acknowledgements** The authors want to thank Neel Krishnaswami for discussions and encouragement.

## A Program comparison with Lucid Synchrone and Zélus

Lets consider the example we give in the introduction, the sum of an alternating signal. The example written in TSD is shown below:

```
let alt =
  let state = ref 1 in
  link state (1 - deref state); deref state
in

let sum = λinp.
  let state = ref 0 in
  link state (inp + deref state); deref state
in

let alt_sum = sum alt in ...
```

The Zélus implementation is remarkably similar:

```
let trans (s, i) = 1 - s
let node alt () = state where
  rec state = 1 -> trans ((pre state), 0)

let plus (x, y) = x + y
let node sum inp = state where
  rec state = 0 -> plus ((pre state), inp)
```

```

let node alt_sum () = sum (alt ())
let node main () = alt_sum ()
```

The key distinction is the Zélus definition of the feedback loop via recursion, versus the TSD definition in which the same loop is created via assignment. Also, TSD dataflow graphs are productive by construction, since cyclic links can only be constructed via cells. So productivity checks are not necessary.

## B Sieve of Eratosthenes in Lucid Synchrone

```
let node first x = v
  where rec v = x fby v

let rec node sieve x =
  let clock filter = (x mod (first x))<> 0
  in merge filter
          (sieve (x when filter))
          (true fby false)

let node from n = o where rec o = n fby (o + 1)

let clock sieve = sieve (from 2)

let node primes () = (from 2) when sieve
```

## C Implementation

The cell is implemented as a reference storing the current value and a thunk representing the dataflow graph. The `graph` type corresponds to a term in the TSD calculus and can either be a dependency `Thunk`, or a `Cell`. There is also a special if-then-else graph `IF_Thunk`, as the behaviour of branching terms is different in the dataflow graph (both branches are evaluated). The `lift` function embeds any standard OCaml term into a `graph` term. Functions `peek`, `step`, `link`, `assign`, `root` work as described earlier. Notice that the `deref` operation is not implemented in order to keep the syntax succinct. As a result, the difference between a cell and a non-cell graph is not known in compile time which could leads to runtime error when one trying to use `link` on these graphs.

Internally, the dataflow graph is maintained using a heterogeneous list of all cells. The `Obj` module is used to overcome OCaml's type restrictions. However, the internal details are inaccessible to the programmer, who may use the module only via the PPX extension `[%dfg t]` which *translates* a *pure dataflow* term t into a graph. By pure dataflow, we mean that $t$ doxes not contain ref, link, assign, root, peek nor step. These functions are not translated and should be used outside the `[%dfg t]` tag. The translation ($\ulcorner - \urcorner_V$) is indexed by a set $V$ of variables bounded in the `[%dfg t]` tag:

$\ulcorner x \urcorner_V$ = x (if $x \notin V$)

```
⌜x⌝_V = Thunk (fun() -> x)} (if x ∈ V)
⌜c⌝_V = Thunk (fun() -> c)
⌜λx.t⌝_V = Thunk (fun() -> fun x -> peek (⌜t⌝_{V∪{x}}))
⌜t u⌝_V = apply ⌜t⌝_V ⌜u⌝_V
⌜if t1 then t2 else t3⌝_V = ite t1 (fun() -> ⌜t2⌝_V)(fun() -> ⌜t3⌝_V)
```

The main helper functions are:

`apply` simulates the application in the TSD calculus,

`ite` implements the semi-strict behaviour of branching,

Their implementations are:

```
let apply t u = Thunk (fun () -> (peek t) (peek u))
let ite b t1 t2 =
  let t1 = t1() in let t2 = t2() in
  IF_Thunk (fun () -> if peek b then t1 else t2)
```

The incremental implementation is similar, except that a graph structure is maintained throughout the execution. Incrementalisation is accomplished by marking certain cells as dirty and only these cells are recomputed in a `step`.

### C.1 Benchmarks

We tested two versions of the code, with and without incrementalisation. The latter is an optimisation in which not all cells are refreshed, but only those that need to. We ran similar (as similar as possible) versions of the same program against Reactive ML, JS Incremental, and Zélus. We chose the programs from the documentations of those programming languages, presumably as representative of their strengths. The examples were:

1. the alternating-sum automata of the introduction,
2. a large cellular automaton,
3. a long-and-thin chain of propagations,
4. a short-and-wide chain of propagations,
5. finite and infinite input response filters,
6. fold-like computations,
7. a simulation of a heating element,
8. a light controller,
9. a map-like computation,
10. a page-rank algorithm,
11. a pendulum simulation,
12. an $n$-body gravitational simulation,
13. and a tree propagation.

For each example we could vary the size of the graph for up to $10^7$ nodes. The benchmark tables are in the Appendix (appendix C.1).

**RML:** As a low-level language with its own compilers we expected RML to be much faster. However, for most examples at least one of the TSD implementations had better or similar performance. RML only outperformed both TSD versions for infinite input response filter and map.

**Incremental JS:** This is a highly-tuned OCaml library used in production by JaneStreet. It does not support cyclic dependencies so most examples could not be implemented in it. For long-and-thin and tree-like chains of propagations incremental TSD outperformed, and for short-and-wide chains plain TSD outperformed.

**Zélus:** For the heater simulation and the light controller TSD was heavily outperformed it. For the pendulum simulation Zélus was slightly faster in very large systems.

The first column represents the size of the dataflow graph. All times expressed in milliseconds.

## D Type Soundness Proof

### D.1 Conventions

In the following sections we will denote certain graph compositions by the rules below which are borrowed from string diagrams.

**Composition** For any graph $G : \Lambda \to (\Gamma \cup \Psi)$ and any graph $E : (\Gamma \cup \Pi) \to \Omega$, the resulting composite graph is denoted by $EG : (\Lambda \cup \Pi) \to (\Psi \cup \Omega)$.

**Parallel composition** For any graph $G : \Lambda \to \Gamma$ and any graph $E : \Pi \to \Omega$, the resulting parallel composite graph is denoted by $E \otimes G : (\Lambda \cup \Pi) \to (\Gamma \cup \Omega)$.

### D.2 Environments

In this section, we identify a special form of graphs, dataflow environments, which corresponds to dataflow graph model. We also prove termination for these graphs in Sec. D.5.

**Definition D.1** (Boxes). There is only one kind of box, the !-box shown below:

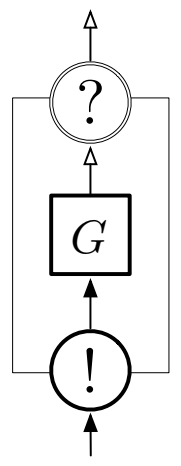

**Definition D.2** (Value dags). A *value dag* is a directed acyclic graph of !-boxes and $C$-nodes.

**Definition D.3** (Term graphs). A *term graph* is a graph that does not have any $\{n\}$-nodes.

**Table 4** Benchmarks against RML and Zélus

**(a)** alternating-sum automata

| size | TSD | inc. TSD | RML | Zélus |
|---|---|---|---|---|
| $10^5$ | 14 | 15 | 33 | 38 |
| $10^6$ | 136 | 146 | 306 | 92 |
| $10^7$ | 835 | 1463 | 2786 | 776 |

**(b)** heater simulation

| size | TSD | inc. TSD | Zélus |
|---|---|---|---|
| $10^4$ | 1 | 1 | 28 |
| $10^5$ | 1 | 2 | 49 |
| $10^6$ | 2 | 1 | 234 |

**(c)** light controller

| size | TSD | inc. TSD | RML | Zélus |
|---|---|---|---|---|
| $10^5$ | 12 | 14 | 71 | 30 |
| $10^6$ | 140 | 138 | 705 | 89 |
| $10^7$ | 1301 | 1292 | 6466 | 648 |

**(d)** pendulum

| size | TSD | inc. TSD | Zélus |
|---|---|---|---|
| $10^5$ | 85 | 191 | 88 |
| $10^6$ | 861 | 1887 | 638 |
| $10^7$ | 8793 | 21 221 | 6464 |

**(e)** cellular automata

| size | TSD | inc. TSD | RML |
|---|---|---|---|
| $10^2$ | 2 | 2 | 8 |
| $10^3$ | 4 | 4 | 897 |
| $10^4$ | 74 | 88 | 1601 |

**(f)** planets

| size | TSD | inc. TSD | RML |
|---|---|---|---|
| 10 | 6 | 6 | 7 |
| $10^2$ | 323 | 336 | 597 |
| $10^3$ | 46 644 | 47 698 | 54 268 |

**(g)** FIR filter

| size | TSD | inc. TSD | RML |
|---|---|---|---|
| $10^2$ | 32 | 226 | 32 |
| $10^3$ | 381 | 3905 | 272 |
| $10^4$ | 11630 | 46323 | 7728 |

**(h)** IIR filter

| size | TSD | inc. TSD | RML |
|---|---|---|---|
| 10 | 116 | 431 | 67 |
| $10^2$ | 1243 | 8351 | 468 |
| $10^3$ | 17 765 | 198 376 | 6465 |

**(i)** page rank

| size | TSD | inc. TSD | RML |
|---|---|---|---|
| 10 | 1 | 1 | 2 |
| $10^2$ | 10 | 2 | 11 |
| $10^3$ | 2970 | 387 | 3060 |

**(j)** map

| size | TSD | inc. TSD | RML |
|---|---|---|---|
| $10^3$ | 2 | 1 | 2 |
| $10^4$ | 34 | 14 | 2 |
| $10^5$ | 462 | 188 | 23 |

**(k)** fold-max

| size | TSD | inc. TSD | RML |
|---|---|---|---|
| $10^2$ | 1 | 2 | 2 |
| $10^3$ | 170 | 2 | 1 |
| $10^4$ | 5381 | 32 | 2 |

**(l)** fold-max 2

| size | TSD | inc. TSD | RML |
|---|---|---|---|
| $10^2$ | 1 | 2 | 2 |
| $10^3$ | 2 | 2 | 2 |
| $10^4$ | 13 | 26 | 2 |

**(m)** fold-sum

| size | TSD | inc. TSD | RML |
|---|---|---|---|
| $10^2$ | 2 | 2 | 2 |
| $10^3$ | 141 | 5 | 1 |
| $10^4$ | 52 868 | 51 | 2 |

**(n)** fold-sum 2

| size | TSD | inc. TSD | RML |
|---|---|---|---|
| $10^2$ | 2 | 1 | 2 |
| $10^3$ | 2 | 2 | 2 |
| $10^4$ | 10 | 26 | 2 |

**Table 5** Benchmarks against JS incremental

**(a)** chain of propagation

| size | inc. TSD | JS Incr. |
|---|---|---|
| $10^3$ | 3 | 5 |
| $10^4$ | 11 | 15 |
| $10^5$ | 167 | 52 |
| $10^6$ | 1796 | 733 |
| $10^7$ | 2243 | 8222 |

**(b)** tree propagation

| size | TSD | inc. TSD | JS Incr. |
|---|---|---|---|
| $10^3$ | 2 | 2 | 5 |
| $10^4$ | 13 | 2 | 7 |
| $10^5$ | 121 | 5 | 15 |
| $10^6$ | 1629 | 28 | 100 |
| $10^7$ | 15 510 | 275 | 998 |

**(c)** field of propagation

| size | TSD | inc. TSD | JS Incr. |
|---|---|---|---|
| $10^3$ | 2 | 1 | 7 |
| $10^4$ | 8 | 3 | 7 |
| $10^5$ | 152 | 12 | 20 |
| $10^6$ | 1683 | 48 | 132 |
| $10^7$ | 17 217 | 384 | 1172 |

**Definition D.4** (Dataflow environment)**.** A dataflow environment is the largest graph connecting (in both direction) to all $\{n\}$-node and consisting of only $\{\$, \text{if}, C, d\}$-nodes satisfying the following conditions:

1. the out-port of any $C$-node is not connected to a $C$-node nor a !-box, and
2. the left out-port of any if-node is not connected to a !-box, and
3. at least one of the out-ports of any \$-nodes is not connected to a !-box.

### D.3 Query and Answer

**Definition D.5** (Answer sets)**.** We define the answer sets any $e : \theta$ as follows:

$$
\begin{aligned}
Ans_{e:Int} &= \{(n, -)\} \\
Ans_{e:Cell} &= \{(n, e)\} \\
Ans_{e:!Int} &= \begin{cases} \{(n, g)\} & \textit{if e is connected to a dataflow environment} \\ \{(n, -)\} & \textit{otherwise} \end{cases} \\
Ans_{e:Unit} &= Ans_{e:!Unit} = \{((), -)\} \\
Ans_{e:\kappa_1 \to \kappa_2} &= Ans_{e:!(\kappa_1 \to \kappa_2)} = \{(\lambda, -)\}
\end{aligned}
$$

**Definition D.6** (Terminating graphs)**.** For any graph $G : \Lambda \to \emptyset$ and any port $e : \kappa \in dom(\Lambda)$, $(G, e) \Downarrow$ iff there exists some $G'$ such that $Init(G, e) \mapsto^* Final(G', e, X)$ for some $X \in Ans_e$.

**Lemma D.1** (Stack extension over transitions)**.** *For any transitions*

$$(G, (e, d, f, S, B), \emptyset) \mapsto (G', (e', d', f', S', B'), \emptyset),$$

*we also have for any* $S_e, B_e$,

$$(G,(e,d,f,SS_e,BB_e),\emptyset) \mapsto (G',(e',d',f',S'S_e,B'B_e),\emptyset)$$

*Proof.* This is proved by case analysis on all transitions. □

**Corollary D.1.1** (Termination extending over stacks). *For any graph* $G : \Lambda \to \emptyset$ *and any port* $e : \kappa \in dom(\Lambda)$ *inside* $G$*, if* $(G,(e,d,f,S,B),\emptyset) \mapsto^* (G',(e',d',f',S',B'),\emptyset)$ *then for any* $S_e, B_e$ *we also have* $(G,(e,d,f,SS_e,BB_e),\emptyset) \mapsto^* (G',(e',d',f',S'S_e,B'B_e),\emptyset)$

### D.4 Valid States

**Definition D.7** (Paths). A path is an ordered list of ports $e_0 ... e_n$ such that if $e_i$, $e_{i+1}$ is the incoming and the outgoing port of same node.

**Definition D.8** (Reachability). A node $X$ is reachable from a port $e$ if there exists a path from $e$ to the incoming port of $X$.

**Definition D.9** (Evaluated in-ports). An in-port is *evaluated* if its reachable graph $K$ is an environment.

**Definition D.10** (Circular requirements). A graph $G$ satisfies the *circular requirements* if all its circular paths satisfy one of the following conditions:

1. the path includes the right and left outgoing port of a $\lambda/\mu$-node and all the ports and nodes in this path are **only** reachable through the respective $\lambda/\mu$-node,
2. the path includes a $\{n\}$-node.

**Definition D.11** (Valid graphs). A graph $G$ is *valid* if it satisfies all the following conditions:

1. any !-nodes, ?-nodes, $\lambda$-nodes and $\mu$-nodes must only appear in the form of boxes,
2. the left out-port of a $\lambda$-node or a $\mu$-node must either be connected by a $C$-node without any in-port or be reachable from its right out-port,
3. the out-port of any $C$-node or ?-node must be evaluated or to a $?/C$-node or the left-in-port of a $\lambda$-node or a $\mu$-node,
4. the out-port of any $\{n\}$-node must be evaluated, and
5. it satisfies the *circular requirements*.

**Definition D.12** (Valid positions). For any graph $G$, a token $\delta = (e,d,f,S,B)$ is said to have a *valid position* if it satisfies all the following conditions:

1. if $e$ is inside a !-box, it must be the out-port of the !-node which has no outer boxes, and
2. $e$ is not the in-port nor out-port of any ?-node.

**Definition D.13** (Valid rewrite flags). For any graph $G$, a token $\delta = (e,d,f,S,B)$ is said to have a *valid rewrite flag* if it satisfies one of the following conditions:

1. if $f = @$ then $e$ must be the in-port of an @-node and $d = \uparrow$,
2. if $f = \text{if}$ then $e$ must be the middle/right out-port of an if-node and $d = \uparrow$,

3. if $f \in \{!, C\}$ then $e$ must be the out-port of the respective node and $d = \uparrow$,
4. if $f = \mu$ then $e$ must be the in-port of a $\mu$-node and $d = \uparrow$,
5. if $f \in \{m, p, r(i), l(i), a(b, i)\}$ then $e$ must be the in-port of the respective node and $d = \downarrow$,
6. if $f \in \{sp, s\}$ then $e$ must be the in-port of an $s$-node with $d = \uparrow$ or $d = \downarrow$ respectively,
7. $f = \square$.

**Definition D.14** (Valid stacks). For any graph $G$ and any path $e_0 \ldots e_n$, an eval token $\delta = (e_n, d, f, S, B)$ is said to have *valid stacks* relative to $p$ if textvalid P' d S B = true where P' is the reverse of $P$ and isValid is defined as follows:

```
isValid :: Path -> Direction -> C-stack -> B-stack -> Boolean
isValid p ↑ (⋆ : S) B = validHist p S B
isValid (e_n : p) ↓ (X : S) B = X ∈ Ans_{e_n} and validHist (e_n : p) S B
isValid _ _ _ _ = false

validHist :: Path -> C-stack -> B-stack -> Boolean
validHist [] [] [] = true
validHist [] _ _ = false
validHist (e_i : p) S B =
  if e_i is the left-output of a #-node then
    S == X : S' and X ∈ Ans_{e'} and validHist p S' B
      where e' is the right-output of the #-node
  else if e_i is the middle-output of an if-node and i < n then
    (S == if_0 : S' or S == if_1 : S') and validHist p S' B
  else if e_i is the right-output of an if-node and i < n then
    S == X : S' and X ∈ Ans_{e'} and validHist p S' B
      where e' is the middle-output of the if-node
  else if e_i is the output of a C-node then
    B == e_{i-1} : B' and validHist p S B'
  else
    validHist p S B
```

**Definition D.15** (Valid graphs relative to paths). For any graph $G$, any path $p = e_0 ... e_n$ and token $(e_n, d, f, X : S, B)$, $G$ is *valid relative to* $p$ if it satisfies the following conditions:

1. if $e_n$ is the out-port of a !-node, $d = \uparrow$ and $f = \square$ then this !-node cannot be connected by a $C$-node,
2. if $e_n$ is the in-port or out-port of a !-node and $d = \downarrow$ then this !-node cannot be connected by a $C$-node,
3. there cannot be two consecutive ports in $p$ being the out-port of a $C$-node except when $d = \uparrow, f = C$ and $e_n$ is the out-port of a $C$-node,
4. $e_n$ must not be only reachable by an $\lambda$-node or a $\mu$-node (that is not inside a box) from its right out-port nor its left in-port,
5. if $d = \downarrow$ then $X = (n, g)$ if and only if $e_n$ is connected to a dataflow environment.

**Definition D.16** (Valid tokens). For any graph $G$, any path $p = e_0 ... e_n$, a token $(e_n, d, f, S, B)$ is *valid* relative to $p$ if it has a *valid position*, a *valid rewrite flag*, *valid stacks* relative to $p$ and $G$ is valid relative to $p$.

**Definition D.17** (Valid prop-tokens). A state $\pi = (G, (e, d, f, S, B), P)$ has *valid prop-tokens* if it satisfies all the following conditions:

1. $P$ can only be non-empty when $f = sp$,
2. every token in $P$ must all be valid relative to a unique non-circular path starting from a $\{n\}$-node.

**Definition D.18** (Evaluated requirements). For any graph $G$, an eval token $\delta = (e, d, f, S, B)$ satisfies the *evaluated requirements* if it satisfies the following conditions:

1. if $d = \downarrow$, $f = \square$ and $e : \kappa$ then $e$ is evaluated,
2. if $e$ is reachable by an if-node from the middle-out-port or the right-out-port, then its left-out-port must be evaluated,
3. if $e$ is reachable by an if-node from the right-out-port and $f = \square$, then its middle-out-port must be evaluated,
4. if $d = \downarrow$ and $e$ is the in-port of an $\omega$-node (unary operations) then its out-port must be evaluated,
5. if $e$ is reachable by an #-node (binary operations) or an @-node from the left-out-port, then its right-out-port must be evaluated,
6. if $d = \downarrow$ and $e$ is the in-port of a #-node then all its out-ports are evaluated, and
7. if $e$ is the in-port of an @-node and $f = @$ then all its out-ports are evaluated.

**Definition D.19** (Valid states). A state $\pi = (G, \delta, P)$ is *valid* if $G$ is valid, $\delta$ is valid relative to a non-circular path starting from an in-port of $G$, $P$ is valid and it satisfies the *evaluated requirements*.

**Lemma D.2** (State validity preservation). *For any valid state $\pi$, if $\pi \mapsto \pi'$ then $\pi$ is also valid.*

*Proof.* This is proved by inspecting all possible transitions one-by-one.

For pass transitions and dummy rewrites, we simply needs to check if the resulting state has valid position, valid rewrite flags valid stacks and valid prop-tokens since the graph does not change.

As for rewrite transitions, the trickiest part is to prove that the *circular requirements* holds, i.e. no arbitrary circular path exists. For example, in the $\lambda$@-rewrite, if the left-in-port of the $\lambda$-node is reachable from its right-out-port, an arbitrary circular path would be form only if its left-in-port is reachable from the right-out-port of the @-node which is impossible because all the ports in the path must be reachable through the $\lambda$-node which means this part of the graph is isolated from the right-out-port of the @-node. The other possible rewrite to create circular path is the $l$-rewrite. However a $l$-rewrite will create a circular path that has a cell in it, as such it conforms to the requirements. Other requirements are checked accordingly. $\square$

**Corollary D.2.1.** *For any valid state $\pi$, if $\pi \mapsto \pi'$ then $\pi'$ is also valid.*

### D.5 Termination of Dataflow Environments

**Proposition D.3** (Termination of dataflow environments). *For any valid dataflow environments $E : \Lambda \to \emptyset$ and any $e \in dom(\Lambda)$, there exists some $X \in Ans_e$ such that $Init(E,e) \mapsto^* Final(E,e,X)$ such that no rewrites occur.*

*Proof.* This can be proved by induction on the maximum length ($n$) of the path from $e$ that stops at any $\{n\}$.

**Base Case ($n = 0$).** This case is not applicable.

**Base Case ($n = 1$).** The only case here would be a single $\{n\}$-node which terminates.

**Inductive Step ($n = k + 1$).** This is proved by case analysis on the possible nodes connected by $e$.

**Sub-Case $e$ is connected to a \$-node, if-node, , $C$-node or $d$-node or !-node.** These cases are similar and thus we will show the proofs for \$-node. Running an execution, we have:

$$
\begin{aligned}
G\$,(e,\uparrow,\square,\star:\square,\square),\emptyset &\mapsto G\$,(e_1,\uparrow,\square,\star:\square,\square),\emptyset && (1)\\
&\mapsto^* G\$,(e_1,\downarrow,\square,X:\square,\square),\emptyset && (2)\\
&\mapsto G\$,(e_2,\uparrow,\square,\star:X:\square,\square),\emptyset && (3)\\
&\mapsto^* G\$,(e_2,\downarrow,\square,Y:X:\square,\square),\emptyset && (4)\\
&\mapsto G\$,(e,\downarrow,\$,(m,g):\square,\square),\emptyset && (5)\\
&\mapsto G\$,(e,\downarrow,\square,(m,g):\square,\square),\emptyset && (6)
\end{aligned}
$$

From (1) to (2) and (3) to (4), since the whole graph is a dataflow environment, therefore the at least one of the out-port must be a dataflow environment as well. On the other hand, if the out-port is connected to a constant !-box then it obviously comes back, there fore we apply the induction hypothesis on the remaining graph and Cor. D.1.1. From (4) to (5) by the definition of dataflow environment, one of the out-port must not be connected to a !-box, as such by *validity*, one of $Y, X$ must be in the form of $(m, g)$. □

### D.6 Determinism

In this section we proved that propagation always terminates and produce unique results and hence the transitions are deterministic up to equivalence of propagation sequences. The intuition is that all the $\{n\}$-nodes are connected to a dataflow environment therefore as we have proved before that dataflow environment always terminates and produces the same graph, thus propagation will also terminate.

**Lemma D.4** (Termination of prop tokens). *For any valid state $(G,(r,\uparrow,sp,S,B),[e,\uparrow,\star:\square,\square])$ where $r$ is the in-port of an s-node, and $e : \kappa$ of a $\{n\}$-node, there exists a sequence of transitions $(G,(r,\uparrow,sp,S,B),[e,\uparrow,\star:\square,\square]) \mapsto^* (G,(r,\uparrow,sp,S,B),[e,\downarrow,X:\square,\square])$.*

*Proof.* Since by *validity*, any $\{n\}$-node will always be connecting to a dataflow environment. Therefore this is simply a consequence of Prop. D.3. □

**Proposition D.5** (Termination of propagation). *For any graph $G$, any possible sequences from $(G,(e,\uparrow,sp,S,B),Init^P)$ will eventually reach $(G,(e,\uparrow,sp,S,B),Final^P(\vec{X}))$ for some $\vec{X}$ where $e$ is the in-port of an s-node.*

*Proof.* This is proved by contradiction. If the sequence does not reach the final state, there are two possibilities: 1) the sequence is infinite or 2) the sequence reaches a state which is not final but cannot proceed. For the first case, at least one of sequences of a prop token is infinite which contradicts Lem. D.4. For the second case, it also means that one of the prop token is stuck which also contradicts Lem. D.4. □

**Lemma D.6** (Diamond property). *For any valid state $(G,\delta,P)$ if $(G,\delta,P) \mapsto (G,\delta,P')$ and $(G,\delta,P) \mapsto (G,\delta,P'')$ then there exists a valid state $\pi$ such that $(G,\delta,P') \mapsto \pi$ and $(G,\delta,P'') \mapsto \pi$.*

*Proof.* Since we have $(G,\delta,P) \mapsto (G,\delta,P')$ and $(G,\delta,P) \mapsto (G,\delta,P'')$, if the states are produced with advancing different prop-tokens, i.e.

$$(G,\delta,\{t_1..t_i..t_j..t_n\}) \mapsto (G,\delta,\{t_1..t_i'..t_j..t_n\})$$

and

$$(G,\delta,\{t_1..t_i..t_j..t_n\}) \mapsto (G,\delta,\{t_1..t_i..t_j'..t_n\})$$

. Then we can join the two sequence into $(G,\delta,\{t_1..t_i'..t_j'..t_n\})$ because the transition does not cause any race conditions. □

**Proposition D.7** (Confluence of propagation). *For any valid state $(G,\delta,P)$, if $(G,\delta,P) \mapsto^* (G,\delta,P')$ and $(G,\delta,P) \mapsto^* (G,\delta,P'')$ then there exists a valid state $(G,\delta,K)$ such that $(G,\delta,P') \mapsto^* (G,\delta,K)$ and $(G,\delta,P'') \mapsto^* (G,\delta,K)$.*

*Proof.* This is a consequence of Lem. D.6. □

**Corollary D.7.1.** *There is exactly one unique final state for any sequence from $(G,\delta,P)$.*

By using this corollary, all the possible propagation sequences from

$$(G,(e,\uparrow,sp,S,B),Init^P)$$

to

$$(G,(e,\uparrow,sp,S,B),Final^P(\vec{X}))$$

form an equivalence class.

**Definition D.20** (Step-propagation transitions). We replace the transitions in Def. 3.5 by the following transitions respectively:

1. $(G,(e,\uparrow,\square,\star:S,B),\emptyset) \mapsto (G,(e,\uparrow,sp,\star:S,B),\emptyset)$

2. $(G,(e,\uparrow,sp,\star:S,B),\emptyset) \mapsto (G'',(e,\downarrow,\square,(b,-):S,B),\emptyset)$ where $G''$ and $b$ are obtained by executing the step-propagation.

**Proposition D.8** (Determinism). *The machine is deterministic by replacing Def. 3.5 with Def. D.20.*

*Proof.* This is a consequence of Prop. D.5 and Prop. D.7. ☐

In the section below, whenever we talk about transitions, unless otherwise specifies, we always mean the replaced set of transitions. As a result the set of prop-tokens will always be empty, thus we will represent the machine state as only $G,\delta$ from now on.

### D.7 Expansion of Contractions

In this section we proved that any graph $G$ behaves the same way as expanding all its outter contractions ($C$-nodes that are not inside a box).

**Definition D.21** (U-simulation). A relation $R$ on states is a $U$-simulation if it satisfies the following conditions:

1. if $\pi_1\ R\ \pi_2$ and a transition $\pi_1 \mapsto \pi_1'$ is possible then there exists two states $\pi_1''$, $\pi_2'$ such that $\pi_1' \mapsto^* \pi_1''$ (possibly no transitions), $\pi_2 \mapsto \pi_2'$ and $\pi_1''\ R\ \pi_2'$,
2. $\pi_1\ R\ \pi_2$ then if there is no possible transition for $\pi_1$ then there is no possible transition for $\pi_2$ as well,
3. if $\pi_1\ R\ \pi_2$, then $\pi_1$ is an ending state if and only if $\pi_2$ is an ending state.

**Lemma D.9.** *For any U-simulation $R$ if $\pi_1\ R\ \pi_2$ and $\pi_2 \mapsto^* \pi_2'$ then there exists a state $\pi_1'$ such that $\pi_1 \mapsto^* \pi_1'$ and $\pi_1'\ R\ \pi_2'$.*

*Proof.* This is proved by induction on the sequence $\pi_2 \mapsto^n \pi_2'$.

**Base Case** $n=0$. Then $\pi_2 = \pi_2'$ and since $\pi_1\ R\ \pi_2$ then we can pick $\pi_1' = \pi_1$ and thus it is true.

**Inductive Case** $n=k+1$. We need to show that there exists some state $\pi_1'$ such that $\pi_1 \mapsto^* \pi_1'$ and $\pi_1'\ R\ \pi_2'$. Suppose there is no possible transition from $\pi_1$ then because $\pi_1\ R\ \pi_2$ thus there should not be any transition from $\pi_2$ as well which contradicts the assumption that $\pi_2 \mapsto^{k+1} \pi_2'$, therefore there must exists a $\pi_1'$ such that $\pi_1 \mapsto \pi_1'$. As a result, because $\pi_1\ R\ \pi_2$ there exists some $\pi_1''$, $\pi_2^*$ such that $\pi_1' \mapsto^* \pi_1''$ such that $\pi_1''\ R\ \pi_2^*$. Since $\mapsto$ is deterministic (Prop. D.8) therefore we have $\pi_2 \mapsto \pi_2^* \mapsto^k \pi_2'$. By using the induction hypothesis, there must exists some $\pi_1^*$ such that $\pi_1'' \mapsto^* \pi_1^*$ and $\pi_1^*\ R\ \pi_2'$. ☐

**Lemma D.10.** *For any U-simulation $R$ if $\pi_1\ R^+\ \pi_2$ and $\pi_2 \mapsto^* \pi_2'$ then there exists a state $\pi_1'$ such that $\pi_1 \mapsto^* \pi_1'$ and $\pi_1'\ R^+\ \pi_2'$.*

*Proof.* This proved by induction on $n \geq 1$ for $\pi_1\ R^n\ \pi_2$.

**Base Case** $n = 1$. This is proved in Lem. D.9.

**Inductive Case** $n = k + 1$. We need to show that if $\pi_1\ R^k \pi_2\ R\ \pi_3$ and $\pi_3 \mapsto^* \pi_3'$ then there exists some $\pi_1'$ such that $\pi_1 \mapsto^* \pi_1'$ and $\pi_1\ R^+\ \pi_1'$. Since $\pi_2\ R\ \pi_3$ by using Lem. D.9, there exists some $\pi_2'$ such that $\pi_2 \mapsto^* \pi_2'$ and $\pi_2'\ R^+\ \pi_3'$. Now by using the induction hypothesis on $k$, there must exists some $\pi_1'$ such that $\pi_1 \mapsto^* \pi_1'$ and $\pi_1'\ R^+\ \pi_2'$. Thus we have $\pi_1'\ R^+\ \pi_3'$. □

**Lemma D.11.** *For any U-simulation R its transitive closure $R^+$ is also a U-simulation.*

*Proof.* We need to show that $R^+$ satisfies the three conditions of a U-simulation.

**Case** (1). We need to show that if $\pi_1\ R^+\ \pi_e$ and a transition $\pi_1 \mapsto \pi_1'$ is possible then there exists two states $\pi_1''$, $\pi_e'$ such that $\pi_1' \mapsto^* \pi_1''$, $\pi_e \mapsto \pi_e'$ and $\pi_1''\ R^+\ \pi_e'$.

Since there must exists a $n \geq 1$ such that $\pi_1\ R^n\ \pi_e$, therefore we can do induction on $n$. When $n = 1$ it is trivially true because $\pi_1\ R\ \pi_e$. When $n = k + 1$ then we have $\pi_1\ R^k\ \pi_2\ R\ \pi_e$. Since $\pi_1\ R^k\ \pi_2$ and $\pi_1 \mapsto \pi_1'$ is possible therefore by using the induction hypothesis, there exists two states $\pi_1''$, $\pi_2'$ such that $\pi_1' \mapsto^* \pi_1''$, $\pi_2 \mapsto \pi_2'$ and $\pi_1''\ R^+\ \pi_2'$. Since $\pi_2\ R\ \pi_e$ and $\pi_2 \mapsto \pi_2'$, therefore these also exists two states $\pi_2''$ and $\pi_e'$ such that $\pi_2' \mapsto^* \pi_2''$, $\pi_e \mapsto \pi_e'$ and $\pi_2''\ R\ \pi_e'$. Then by using Lem. D.10, there must exists some state $\pi_1'''$ such that $\pi_1'' \mapsto^* \pi_1'''$ and $\pi_1'''\ R\ \pi_2''$. Thus $\pi_1'''\ R^+\ \pi_e'$.

**Case** (2). We need to show that if $\pi_1\ R^+\ \pi_e$ and there is no possible transition from $\pi_1$ then so is $\pi_e$.

Since there must exists a $n \geq 1$ such that

$$\pi_1\ R\ \pi_2\ R\ \ldots\ R\ \pi_{n-1}\ R\ \pi_e$$

therefore we can do induction on $n$. When $n = 1$ we have $\pi_1\ R\ \pi_e$ and thus $\pi_e$ has no possible transition. When $n = k + 1$ we have $\pi_1\ R\ \pi_2\ R\ \ldots\ R\ \pi_k\ R\ \pi_e$. Since $\pi_1\ R\ \pi_2$ means that $\pi_2$ has no possible transitions and hence by the induction hypothesis this is true.

**Case** (3). We need to show that if $\pi_1\ R^+\ \pi_e$ then $\pi_0$ is an ending state if and only if $\pi_e$ is an ending state.

Since there must exists a $n \geq 1$ such that

$$\pi_1\ R\ \pi_2\ R\ \ldots\ R\ \pi_{n-1}\ R\ \pi_e$$

, therefore we can do induction on $n$. When $n = 1$ it is trivially true because $\pi_1\ R\ \pi_e$. When $n = k + 1$ we have $\pi_1\ R\ \pi_2\ R\ \ldots\ R\ \pi_k\ R\ \pi_e$. a) $\pi_1$ is an ending state implies $\pi_e$ is an ending state: Since $\pi_1\ R\ \pi_2$ means that $\pi_2$ is also an ending state therefore by the induction hypothesis, $\pi_e$ is also an ending state. The other direction is similar. □

**Lemma D.12.** *For any states $\pi_1$, $\pi_2$ and an U-simulation R such that $\pi_1\ R\ \pi_2$, then $\pi_1 \mapsto^* \pi_1^e \iff \pi_2 \mapsto^* \pi_2^e$ for some ending states $\pi_1^e, \pi_2^e$.*

*Proof.* For the direction $\pi_1 \mapsto^* \pi_1^e \implies \pi_2 \mapsto^* \pi_2^e$, it can be proved by induction on the number of sequence $\pi_1 \mapsto^n \pi_1^e$.

**Base Case** $n = 0$. Since $\pi_1 = \pi_1^e$ and $\pi_1 \; R \; \pi_2$, therefore $\pi_2$ is an ending state.

**Inductive Step** $n = k + 1$. We need to show that if $\pi_1 \mapsto^{k+1} \pi_1^e$ then there must exists an ending state $\pi_2^e$ such that $\pi_2 \mapsto^* \pi_2^e$.

We can factorise the sequence into $\pi_1 \mapsto \pi_1' \mapsto^k \pi_1^e$. By the definition of U-simulation, there must exists two states $\pi_1''$ and $\pi_2'$ such that $\pi_1' \mapsto^* \pi_1''$, $\pi_2 \mapsto \pi_2'$ and $\pi_1'' \; R \; \pi_2'$. Therefore by determinism, either 1) $\pi_1''$ is an ending state which then by definition $\pi_2'$ has to be an ending state or 2) $\pi_1'' \mapsto^i \pi_1^e$ where $i \leq k$ and thus by induction hypothesis, there must exists some state $\pi_2^e$ such that $\pi_2' \mapsto^* \pi_2^e$.

For the other direction $\pi_1 \mapsto^* \pi_1^e \impliedby \pi_2 \mapsto^* \pi_2^e$, it is proved in Lem. D.9. □

**Definition D.22** (Replacement of contractions)**.** We define a relation $\prec$ (replace by) on graphs where $\prec = \prec_m \cup \prec_c \cup \prec_i$

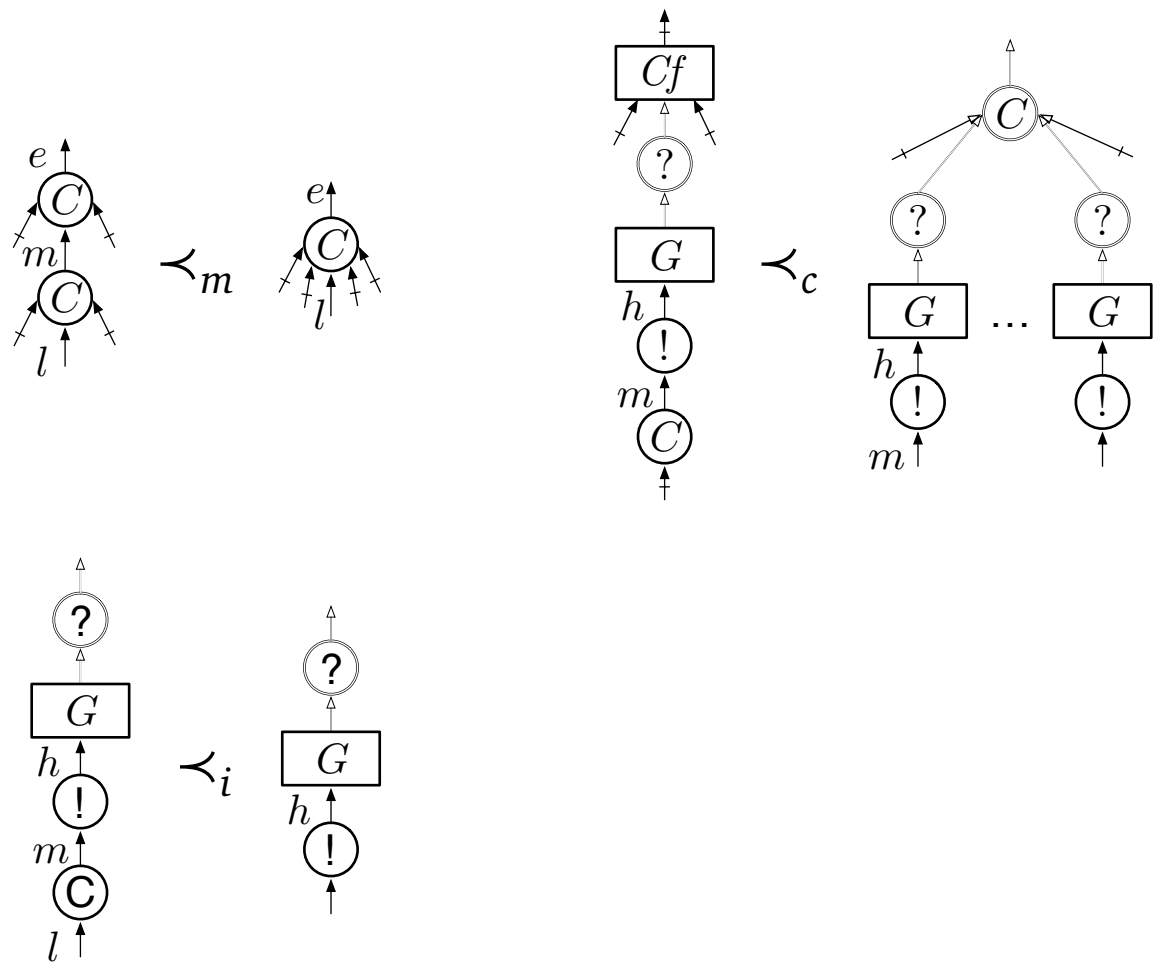

where $Cf$ is the largest forest of $C$-nodes connected, and the $C$-node on the left-hand-side of $\prec_c$ has more than one in-ports.

**Definition D.23.** We define a binary relation $\hat{\prec}$ on *valid* states such that for any graph context $\mathcal{G}[\square]$, $(\mathcal{G}[G_1], \delta_1) \; \hat{\prec} \; (\mathcal{G}[G_2], \delta_2)$ if $G_1 \prec G_2$ where the bottom $C$-node is not connected by another $C$-node, or $G_1 = G_2$, $\delta_1 = \delta_2$, the position of the main token is in $\mathcal{G}$. Note that if $Cf$ is empty, then the only in-ports for $G1$ would be the in-ports of the $C$-node.

**Lemma D.13.** *$\hat{\prec}$ is a U-simulation.*

*Proof.* This is proved by analysing all the possible transitions one-by-one. In some of the following cases, we will only show the part of the graph that matters.

**Case when token is inside** $G$ **excluding the border**. If it is a pass transition or a dummy rewrite, then we also have the exact same transition on the right-hand-side.

If it is a rewrite transition then it depends on whether it involves any of the holes. If not then it is trivial. If yes, then it must be one of the following cases (other cases are disproved by *validity*): (1)

$($ 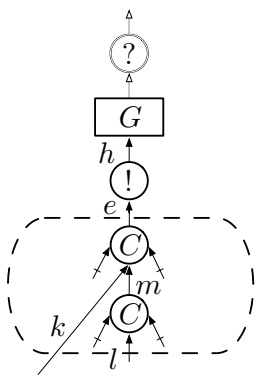 $, h, \uparrow, !, S, k : B)$

$\mapsto ($ 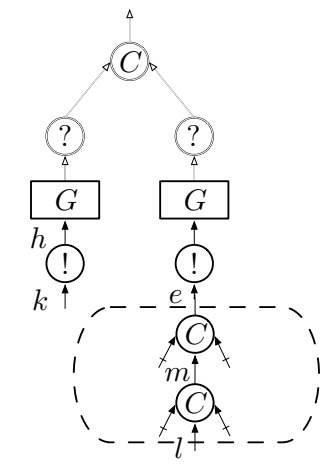 $, h, \uparrow, !, S, B)$

*vs*

$($ 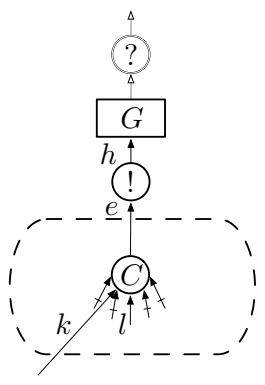 $, h, \uparrow, !, S, k : B)$

$\mapsto ($ 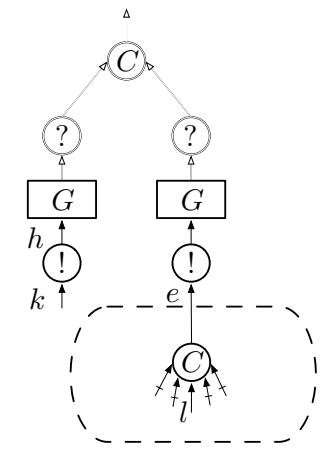 $, h, \uparrow, !, S, B)$

(2) we have $C[CC]$ vs $C[C]$ where the token is at the top of the highest $C$-node. In this case by validity, the current execution path does not contain any in-ports or out-ports of from the hole.

And (3) we have $?G![(Cf)(?G!)C]$ vs $?G![C(?\vec{G}!)]$. In this case copying the upper !-box using a $C$-node in $Cf$ will still result in a forest of $C$-nodes and so it is fine.

**Case when token is at the border with $G_1 \prec_m G_2$.** We proved this by analysing all the possible cases.

**Sub-Cases when token is at an in-port with $d = \uparrow$ and $f \neq \square$.** These cases induce rewrites in the ambient graph that do not affects the hole and thus they have one-to-one correspondence.

**Sub-Case when token is at an in-port with $d = \uparrow$ and $f = \square$.** Since $l$ cannot be connected to another $C$-node therefore we have only two cases: (1)

$(\mathscr{G}[$ 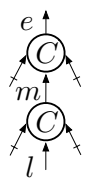 $], l, \uparrow, \square, S, B) \mapsto^4 (\mathscr{G}[$ 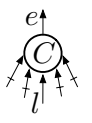 $], e, \uparrow, C, S, l : B)$

$$vs$$

$(\mathscr{G}[$ 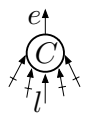 $], l, \uparrow, \square, S, B) \mapsto (\mathscr{G}[$ 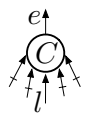 $], e, \uparrow, C, S, l : B)$

and (2)

$($ 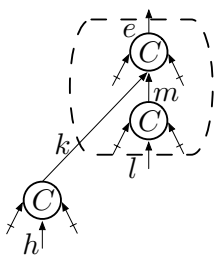 $, k, \uparrow, \square, S, h : B) \mapsto^4 ($ 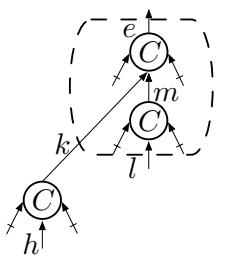 $, e, \uparrow, C, S, k : h : B)$

$$vs$$

$($ 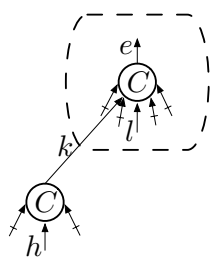 $, l, \uparrow, \square, S, h : B) \mapsto ($ 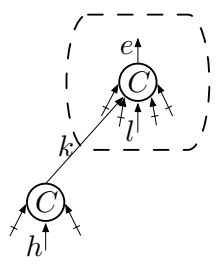 $, e, \uparrow, C, S, k : h : B)$

**Sub-Case when token is at an in-port and** $d = \downarrow$. By validity, the only possible case is when the token is at an in-port of the upper $C$-node that is not $m$ and a transition takes the token away from the border to $G$ which will be the same transition for the right-hand-side.

**Sub-Case when token is at an out-port and** $d = \uparrow$. It is similar to the case when token is at the in-port with $d = \uparrow$. The possible transition would either be a rewrite of the hole or if the upper $C$-node is connected by another $C$-node then it would be a rewrite of those two. It can also be the case that $f = \square$ and any possible transitions will take the token back to the context.

**Sub-Case when token is at an out-port and** $d = \downarrow$. By validity, $B = j : B'$ such that $j \neq m$, as such the token goes back to an in-port that is not from the lower $C$-node.

**Case when the token is at the border of** $G_1 \prec_c G_2$. We proved this by analysing all possible cases.

**Sub-Case when token is at an in-port and** $d = \uparrow$. We have (1) the token is at one of the in-ports of the bottom $C$-nodes, and (2) the token is at one of the in-ports of $Cf$. These two cases are proved similarly as the $\prec_c$-cases.

**Sub-Case when the token is at the bottom and** $d = \downarrow$. By validity the token cannot be at one of the in-ports of the $C$-node, so the only case is when the token is at one of the in-ports of $Cf$. Any possible transitions will take the token back into the ambient graph and the right-hand-side will have the same transition.

**Sub-Case when the token is at the top.** By validity, the possible cases would be that the path does not incudes the !-box nor the $C$-node. So it would either be a rewrites inside the $Cf$ or a transition that takes the token back to the ambient graph.

**Case when token is at the border of** $G_1 \prec_e G_2$. The only possible cases are when the token is at the bottom with $d = \uparrow$ and it follows the same idea as previous cases. □

**Proposition D.14.** *For any graphs* $H_1, H_2 : \Lambda \to \emptyset$*, if* $H_1 \prec^+ H_2$ *then for any* $e \in dom(\Lambda)$*,* $(H_1, e) \Downarrow \iff (H_2, e) \Downarrow$ *where* $\prec^+$ *is the transitive closure of* $\prec$*.*

*Proof.* Since $Init(H_1, e) \mathrel{\hat{\prec}}^+ Init(H_2, e)$ therefore this is simply a consequence of Lem. D.13, Lem. D.11 and Lem. D.12. □

### D.8 Re-evaluation

**Definition D.24** (Stable sequences)**.** A sequence is *stable* if it does not contain any actual rewrite transitions.

**Lemma D.15** (Factorisation of stable sequences)**.** *For any graph* $G : (\{e : \kappa\} \cup \Lambda) \to \emptyset$*, if we have a stable sequence* $Init(G, e) \mapsto^* (G, e', \downarrow, \square, X : S, B)$ *then it can be factorised into* $Init(G, e) \mapsto^* (G, e', \uparrow, \square, \star : S, B) \mapsto^* (G, e', \downarrow, \square, X : S, B)$*.*

*Proof.* This is proved by induction on the number of transitions $Init(G, e) \mapsto^n (G, e', \downarrow, \square, X : S, B)$.

Base Case $n = 0$: This is impossible since the start state must have $\uparrow$.

Inductive Case $n = k + 1$: Since we have $Init(G, e) \mapsto^n (G, \delta) \mapsto (G, e', \downarrow, \square, X : S, B)$, this can be proved by case analysis on the last transition. We will show the proof for $(\mathscr{G}[\$], e', \downarrow, \$, (n, g) : S, B) \mapsto (\mathscr{G}[\$], e', \downarrow, \square, (n, g) : S, B)$ where $e'$ is the in-port of the \$-node. Since the machine is deterministic we can rewind the sequence to get

$$\begin{aligned} & Init(\mathscr{G}[\$], e) \\ \mapsto^n\ & (\mathscr{G}[\$], e_l, \downarrow, \square, (n_2, x) : (n_1, y) : S, B) \\ \mapsto\ & (\mathscr{G}[\$], e', \downarrow, \$, (n, g) : S, B) \end{aligned}$$

Now by using induction hypothesis, we also have that

$$\begin{aligned} & Init(\mathscr{G}[\$], e) \\ \mapsto^*\ & (\mathscr{G}[\$], e_r, \downarrow, \square, (n_1, -) : S, B) \\ \mapsto\ & (\mathscr{G}[\$], e_l, \uparrow, \square, \star : (n_1, -) : S, B) \\ \mapsto^*\ & (\mathscr{G}[\$], e_l, \downarrow, \square, (n_2, x) : (n_1, y) : S, B) \\ \mapsto\ & (\mathscr{G}[\$], e', \downarrow, \$, (n, g) : S, B) \end{aligned}$$

Continuing the process we will get

$$\begin{aligned}
& \mathit{Init}(\mathcal{G}[\$], e) \\
\mapsto^* \; & (\mathcal{G}[\$], e', \uparrow, \square, \star : S, B) \\
\mapsto \; & (\mathcal{G}[\$], e_r, \uparrow, \square, \star : S, B) \\
\mapsto^* \; & (\mathcal{G}[\$], e_r, \downarrow, \square, (n_1, -) : S, B) \\
\mapsto \; & (\mathcal{G}[\$], e_l, \uparrow, \square, \star : (n_1, -) : S, B) \\
\mapsto^* \; & (\mathcal{G}[\$], e_l, \downarrow, \square, (n_2, x) : (n_1, y) : S, B) \\
\mapsto \; & (\mathcal{G}[\$], e', \downarrow, \$, (n, g) : S, B)
\end{aligned}$$

□

**Lemma D.16.** *For any graph $G : (\{e : \kappa\} \cup \Lambda) \to \emptyset$, if $\mathit{Init}(G, e) \mapsto^* (G, \delta) \mapsto^R (G', \delta')$ where the first part is stable, then we also have a stable sequence $\mathit{Init}(G', e) \mapsto^* (G', \delta')$.*

*Proof.* This is proved by case analysis on every actual rewrite transitions $(G, \delta) \mapsto^R (G', \delta')$.

**Case $p$-rewrite, $r$-rewrite, $m$-rewrite, $if$-rewrite, \$-rewrite, $l$-rewrite, $a$-rewrite, @-rewrite, $C$-rewrite, $!C$-rewrites, $\mu$-rewrite**: These cases are similar and thus we will show the proof for \$-rewrite. Let $G = \mathcal{G}[\$]$ where $\mathcal{G}$ is the surround context, then we have that

$$\begin{aligned}
& \mathit{Init}(\mathcal{G}[\$], e) \\
\mapsto^* \; & (\mathcal{G}[\$], e', \downarrow, \$, (n, -) : S, B) \\
\mapsto \; & (\mathcal{G}[C_0 \otimes (n!) \otimes C_0], e', \downarrow, \square, (n, -) : S, B)
\end{aligned}$$

where $e'$ is the in-port of the \$-node. Rewinding the sequence and by using Lem. D.15, we have that

$$\begin{aligned}
& \mathit{Init}(\mathcal{G}[\$], e) \\
\mapsto^* \; & (\mathcal{G}[\$], e', \uparrow, \square, \star : S, B) \\
\mapsto \; & (\mathcal{G}[\$], e_r, \uparrow, \square, \star : S, B) \\
\mapsto^* \; & (\mathcal{G}[\$], e_r, \downarrow, \square, (n_1, -) : S, B) \\
\mapsto \; & (\mathcal{G}[\$], e_l, \uparrow, \square, \star : (n_1, -) : S, B) \\
\mapsto^* \; & (\mathcal{G}[\$], e_l, \downarrow, \square, (n_2, -) : (n_1, -) : S, B) \\
\mapsto \; & (\mathcal{G}[\$], e', \downarrow, \$, (n, -) : S, B) \\
\mapsto \; & (\mathcal{G}[C_0 \otimes (n!) \otimes C_0], e', \downarrow, \square, (n, -) : S, B)
\end{aligned}$$

where $e_r, e_l$ are the right-out-port and left-out-port of the \$-node. Since replacing the \$-subgraph with a constant !-box does not affect the first part of the sequence, therefore we can construct the following sequence

$$
\begin{aligned}
& \mathit{Init}(\mathscr{G}[C_0 \otimes (n!) \otimes C_0], e) \\
\mapsto^* \; & (\mathscr{G}[C_0 \otimes (n!) \otimes C_0], e', \uparrow, \square, \star : S, B) \\
\mapsto \; & (\mathscr{G}[C_0 \otimes (n!) \otimes C_0], e'', \uparrow, \square, \star : S, B) \\
\mapsto \; & (\mathscr{G}[C_0 \otimes (n!) \otimes C_0], e'', \downarrow, \square, (n, -) : S, B) \\
\mapsto \; & (\mathscr{G}[C_0 \otimes (n!) \otimes C_0], e', \downarrow, \square, (n, -) : S, B)
\end{aligned}
$$

where $e''$ is the in-port to the $n$-node.

**Case $\mapsto^P$-rewrite**: The proof is similar. We first have that

$$\mathit{Init}(\mathscr{G}[s], e) \mapsto^* (\mathscr{G}[s], e', \uparrow, sp, \star : S, B) \mapsto^P (G', e', \downarrow, \square, (b, -) : S, B)$$

then we want to construct a sequence

$$\mathit{Init}(G', e) \mapsto^* (G', e', \uparrow, sp, \star : S, B) \mapsto^* (G', e', \downarrow, \square, (b, -) : S, B)$$

, suppose the first half of the second sequence does not lead to $G'$ which means that there exists some rewrite transitions. However since the difference between $G$ and $G'$ is that the content of some cells are changed. This would mean that this transitions will also happen in the original sequence which draws a contradiction. □

**Proposition D.17.** *For any graph $G : \{e : \kappa\} \cup \Lambda \to \emptyset$, if $\mathit{Init}(G, e) \mapsto^* (G', \delta)$ for some $G', \delta$, then we also have that $\mathit{Init}(G', e) \mapsto^* (G', \delta)$.*

*Proof.* Since the sequence $\mathit{Init}(G, e) \mapsto^* (G', \delta)$ is finite, therefore this can be proved by induction on the number ($n$) of actual rewrite transitions in this sequence.

**Base Case** $n = 0$: this means that there are no actual rewrites in the sequence and therefore $G = G'$.

**Inductive Step** $n = k + 1$. First we factorise the sequence into

$$\mathit{Init}(G, e) \mapsto^* (G, \delta') \mapsto^R (G'', \delta'') \mapsto^* (G', \delta)$$

where $\mapsto^R$ denotes an actual rewrite transition.

By using Lem. D.16 on $\mathit{Init}(G, e) \mapsto^* (G, \delta') \mapsto^R (G'', \delta'')$, we have that $\mathit{Init}(G'', e) \mapsto^* (G'', \delta'')$. Since the machine is deterministic we have $\mathit{Init}(G'', e) \mapsto^* (G'', \delta'') \mapsto^* (G', \delta)$ with $k$ rewrites transitions. Therefore by using the induction hypothesis on $k$, we have that $\mathit{Init}(G', e) \mapsto^* (G', \delta)$. □

### D.9 Termination of Recursion-free Programs

In this section, we prove that programs without recursion always terminate. The proof of termination relies on a unary logical relation, also known as a "logical predicate".

**Definition D.25** (Logical predicate for terminating graphs)**.** For any term graph $G : \{e : \kappa\} \to \Gamma$, the logical predicate for terminating graphs $T_\kappa$ is inductively defined on $\kappa$ as follows:

1. $\kappa \in \{!Int, !Unit, Cell\}$: $G \in T_\kappa$ iff for any dataflow environment $E : \Lambda \to \emptyset$ such that $\Gamma \subset \Lambda$, $(EG, e) \Downarrow$.
2. $\kappa = !(\kappa_1 \multimap \kappa_2)$: $G \in T_\kappa$ if
   a. for any dataflow environment $E : \Lambda \to \emptyset$ such that $\Gamma \subset \Lambda$, $(EG, e) \Downarrow$, and
   b. for any term graph $H : \{e_1 : \kappa_1\} \to \Gamma'$ such that $H \in T_{\kappa_1}$, $(G \otimes H)@ \in T_{\kappa_2}$, and
   c. for any for any dataflow environment $E : \Lambda \to \emptyset$ such that $\Gamma \subset \Lambda$ if there exists some $K, X$ such that $Init(EG, e) \mapsto Final(K, e, X)$ and that $K$ can be decomposed into $E'G'$ then for any term graph $H : \{e_1 : \kappa_1\} \to \Gamma'$ such that $H \in T_{\kappa_1}$, $(G' \otimes H)@ \in T_{\kappa_2}$.

**Proposition D.18.** *For any term graphs $G_1, G_2 : \{e : \kappa\} \to \Gamma$, $G_1 \prec^+ G_2$ implies that $G_1 \in T_\kappa \iff G_2 \in T_\kappa$.*

*Proof.* This is proved by induction on $\kappa$.

**Base Case $\kappa$ is a ground type**. To show that $G_1 \in T_\kappa \implies G_2 \in T_\kappa$ we need to show that for any dataflow environment $E$, $EG_2 \Downarrow$. Since $(EG_1, e) \Downarrow$ and $EG_1 \prec^+ EG_2$, therefore by using Lem. D.14, $(EG_2, e) \Downarrow$.

**Induction Step** $\kappa = !(\kappa_1 \multimap \kappa_2)$. To show $G_1 \in T_\kappa \implies G_2 \in T_\kappa$ we need to show that (a) $(G_2, e) \Downarrow$, (b) for any term graph $H : \{e_1 : \kappa_1\} \to \Gamma'$ such that $H \in T_{\kappa_1}$, $(G \otimes H)@ \in T_{\kappa_2}$, and (c) for any dataflow environment $E : \Lambda \to \emptyset$ such that $\Gamma \subset \Lambda$ if there exists some $K, X$ such that $Init(EG_2, e) \mapsto Final(K, e, X)$ and that $K$ can be decomposed into $E'G_2'$ then for any term graph $H : \{e_1 : \kappa_1\} \to \Gamma'$ such that $H \in T_{\kappa_1}$, $(G_2' \otimes H)@ \in T_{\kappa_2}$. (a) can be proved by using similar technique as the base case. To show (b) we simply apply the induction hypothesis on $\kappa_2$. The other direction can be shown by similar technique. To show (c) we run a sequence from $Init(EG_1, e)$ to get the resulting graph $EG_1'$ such that $G_1' \prec^+ G_2'$ because of Lem. D.9. Since $G_1 \in T_\kappa$ thus $(G_1' \otimes H)@ \in T_{\kappa_2}$, therefore by induction hypothesis $(G_2' \otimes H)@ \in T_{\kappa_2}$. □

**Definition D.26** (Value trees). A *value tree* is a tree of !-boxes.

**Lemma D.19.** *For any value tree $V : \{e_1 : \kappa\} \to \Lambda$, $V\vec{C_2} \prec^+ \vec{C_2}(V \otimes V)$.*

*Proof.* This is proved by induction on the maximum height $h$ of the value tree.

**Base Case** $h = 0$. This is true as $\vec{C_2} = \vec{C_2}$.

**Base Case** $h = 1$. This is true as $V\vec{C_2} \prec_c \vec{C_2}(V \otimes V)$.

**Induction Step** $h = k + 1$. We need to show that $(G!)V\vec{C_2} \prec^+ \vec{C_2}((G!)V \otimes (G!)V)$ where the maximum height of $V$ is $h$ and $(G!)$ is a value box. By induction hypothesis we have $(G!)V\vec{C_2} \prec^+ (G!)\vec{C_2}(V \otimes V) \prec_c \vec{C_2}((G!)V \otimes (G!)V)$. □

**Lemma D.20.** *For any set of value tress $\vec{V} = V_1 : \{e_1 : \kappa_1\} \to \Lambda_1, \ldots, V_m : \{e_m : \kappa_m\} \to \Lambda_m$, $\vec{V}\vec{C_2} \prec^+ \vec{C_2}(\vec{V} \otimes \vec{V})$.*

*Proof.* This is proved by induction on $m$ using Lem. D.19. □

**Lemma D.21.** *For any value dag $V : \{e : \kappa\} \to \Gamma$, there exists some value tree $V'$ such that $V \prec^{+} \vec{C}V'$.*

*Proof.* This is proved by induction on the maximum height of $V$ using similar technique as Lem. D.19. ☐

**Lemma D.22.** *For any dataflow environment $E : \Lambda \to \emptyset$ and any $\vec{C} : \Omega' \to \Omega$ such that $\Omega \subset \Lambda$, there exists another dataflow environment $E' : \Lambda' \to \emptyset$ such that $E\vec{C} \prec E'$ or $E\vec{C}$ is a dataflow environment it self.*

*Proof.* When a $C$-node in $\vec{C}$ is connected to a $\$, \text{if}, \{n\}$-node then it is obvious that the resulting graph is a dataflow environment. When a $C$-node is connected by another $C$-node, the two will be replaced by a single $C$-node and the resulting graph would still be a dataflow environment. ☐

**Lemma D.23.** *For any term graph $G : \{e : \kappa\} \to \Gamma$ and any set of parallel $C$-nodes $\vec{C} : \Gamma \to \omega$,*

$$G \in T_\kappa \implies \vec{C}G \in T_\kappa \tag{1}$$

$$\vec{C}G \in T_\kappa \implies G \in T_\kappa \tag{2}$$

*where $\vec{C}G : \{e : \kappa\} \to \omega$.*

*Proof.* This is proved by induction on $\kappa$.

**Base Case $\kappa$ is a ground type**. To show (1) we need to show that for any dataflow environments $E$, $(E\vec{C}G, e) \Downarrow$. First by Lem. D.22 there exists a $E'$ such that $E\vec{C}G \prec^{+} E'G$. Since $E'G \Downarrow$ (by assumption) therefore by using Lem. D.14, $(E\vec{C}G, e) \Downarrow$. (2) can be shown by similar technique.

**Induction Step $\kappa = !(\kappa_1 \multimap \kappa_2)$**. To show (1) we need to show that (a) for any dataflow environment $E$, $(E\vec{C}G, e) \Downarrow$, (b) for any term graph $H : \{e_1 : \kappa_1\} \to \Gamma'$ such that $H \in T_{\kappa_1}$, $(\vec{C}G \otimes H)@ \in T_{\kappa_2}$ and (c) for any dataflow environment $E : \Lambda \to \emptyset$ such that $\Gamma \subset \Lambda$ if there exists some $K, X$ such that $\mathit{Init}(E\vec{C}G, e) \mapsto \mathit{Final}(K, e, X)$ and that $K$ can be decomposed into $E'G'$ then for any term graph $H : \{e_1 : \kappa_1\} \to \Gamma'$ such that $H \in T_{\kappa_1}$, $(G' \otimes H)@ \in T_{\kappa_2}$. (a) can be proved by using similar technique as the base case. To show (b) we simply apply the induction hypothesis on $\kappa_2$. For (c) since $E\vec{C} \prec^{+} F$ for some $F$, we run a sequence from $\mathit{Init}(FG, e)$ to get $F'G'$ since $G \in T_\kappa$ therefore it is true (2) can be shown by similar technique. ☐

**Lemma D.24.** *For any term graph $G : \{e : \kappa\} \to \Gamma$ where $\kappa = !(\kappa_a \multimap \kappa_b)$ if for any $\vec{V}$ where $V_i \in T_{\kappa_i}$, $\vec{V}G \in T_\kappa$, then for any $\vec{V}$ where $V_i \in T_{\kappa_i}$, $\vec{V}G@ \in T_{\kappa'}$.*

*Proof.* We need to show that for any $\vec{V}$ where $V_i \in T_{\kappa_i}$, $\vec{V}G@ \in T_{\kappa'}$. Since by assumption we have $\vec{V}G \in T_\kappa$ therefore by the definition of logical predicate and taking $H$ to be the empty graph, we have that $\vec{V}G@ \in T_{\kappa'}$. ☐

**Lemma D.25** (Backward reasoning). *For any term graph $G : \{e : \kappa\} \to \Gamma$, if there exists a graph $G' : \{e : \kappa\} \to \Gamma$ such that (I) for any dataflow environment $E : \Lambda \to \emptyset$ such that $\Gamma \subset \Lambda$, Init$(EG, e) \mapsto^* (E'\vec{V}G', \delta)$ for some $\vec{V}$ and $\delta$ and (II) for any $\vec{V}$ such that $V_i \in T_{\kappa_i}$, $\vec{V}G' \in T_\kappa$, then $G \in T\kappa$.*

*Proof.* This is proved by induction on $\kappa$.

**Base Case $\kappa$ is a ground type**. Then we only need to prove that it terminates. Assuming a dataflow environmrnt $E$, because of the assumption of $G$, we would have $Init(EG, e) \mapsto *(E'\vec{V}G', \delta)$. Also by assumption we have that $\vec{V}G' \in T_\kappa$ and thus $(E'\vec{V}G', e) \Downarrow$. Since the machine is deterministic (Prop. D.8), and by using Lem. D.17, we have $Init(EG, e) \mapsto *(E'\vec{V}G', \delta) \mapsto^* Final(E''K, e, X)$ and thus this case is true.

**Inductive Case $\kappa = !(\kappa_1 \multimap \kappa_2)$**. To prove (a), it follows the same idea as the base case. As for (b), we need to prove that for any term graph $H : \{e_1 : \kappa_1\} \to \Gamma'$ such that $H \in T_{\kappa_1}$, $(G \otimes H)@ \in T_{\kappa_2}$.

Assume a dataflow environment $E$, we would have $Init(E(G \otimes H)@, e_2) \mapsto^* (E'(\vec{V}G' \otimes H')@, \delta)$ because of the assumptions such that $H' \in T_{\kappa_1}$. Notice that for any changes to $E$, we will always run to a graph in the form of $(X'(\vec{Y}G' \otimes X')@$. This satisfies the assumption (II) of the IH if we take $G'$ to be $G'@$. Now by using Lem. D.24, we also have condition (II) of the IH. Therefore we can now apply the IH on $\kappa_2$ and finally get $(G \otimes H)@ \in T_{\kappa_2}$. (c) is obvious because any dataflow environment will take $G$ to $\vec{V}G'$ which already satisfies the logical predicate by assumption. □

**Lemma D.26** (Fundamental lemma). *For any well-typed term $\Gamma \vdash t : \tau$ which is recursion-free, let $G : \{e : \kappa\} \to \Gamma = [[\Gamma \vdash t : \tau]]$ any set of recursion-free value trees $\vec{V} = V_1 : \{e_1 : \kappa_1\} \to \Lambda_1, \ldots, V_m : \{e_m : \kappa_m\} \to \Lambda_m$ such that $V_i \in T_{\kappa_i}$,*

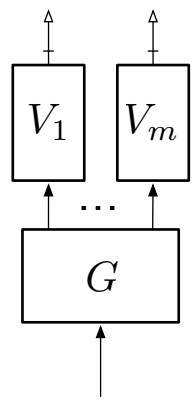

$\in T_\kappa$

*where $\vec{V}G : \{e : \kappa\} \to \Lambda$ and $\Lambda = \bigcup \Lambda_i$.*

*Proof.* This is proved by induction on the type derivation $\Gamma \vdash t : \tau$.

**Base Case** $\Gamma \vdash x : \tau$ which means that $G$ is empty. Since $V_i \in T_{k_i}$ therefore it is true.

**Base Case** $\Gamma \vdash n : Int$. In this case since $\vec{V}$ is empty we only need to show that for any dataflow environment $E : \Lambda \to \emptyset$ such that $\Gamma \subset \Lambda$, $(EG, e) \Downarrow$. The proof is straight-forward.

**Inductive Step** $\Gamma \vdash \lambda x.t : \tau_1 \to \tau_2$. We need to show that (a) for any dataflow environment $E : \Lambda \to \emptyset$ such that $\Gamma \subset \Lambda$, $(E\vec{V}G, e) \Downarrow$, (b) for any term graph $H : \{e_1 : \kappa_1\} \to \Gamma'$ such that $H \in T_{\kappa_1}$, $(\vec{V}G \otimes H)@ \in T_{\kappa_2}$ and (c) for any dataflow environment $E$ :

$\Lambda \to \emptyset$ such that $\Gamma \subset \Lambda$ if there exists some $K, X$ such that $Init(E\vec{V}G, e) \mapsto Final(K, e, X)$ and that $K$ can be decomposed into $E'G'$ then for any term graph $H : \{e_1 : \kappa_1\} \to \Gamma'$ such that $H \in T_{\kappa_1}$, $(G' \otimes H)@ \in T_{\kappa_2}$.

The proof of (a) is straight-forward. In order to show (b), we assume a dataflow environment $E : \Lambda \to \emptyset$ such that $\Gamma \cup \Gamma' \subset \Lambda$ and run a sequence from $Init(E(\vec{V}G \otimes H)@, e_2)$:

$$
\begin{aligned}
&E(\vec{V}G \otimes H)@, (e_2, \uparrow, \square, \star : \square, \square) && (1)\\
\mapsto &E(\vec{V}G \otimes H)@, (e_1, \uparrow, \square, \star : \square, \square) && (2)\\
\mapsto^* &E'(\vec{V}G \otimes H')@, (e_1, \downarrow, \square, X : \square, \square) && (3)\\
\mapsto &E'(\vec{V}G \otimes H')@, (e, \uparrow, \square, \star : \square, \square) && (4)\\
\mapsto &E'(\vec{V}G \otimes H')@, (a, \uparrow, !, \star : \square, \square) && (5)\\
\mapsto &E'(\vec{V}G \otimes H')@, (a, \uparrow, \square, \star : \square, \square) && (6)\\
\mapsto &E'(\vec{V}G \otimes H')@, (a, \downarrow, \square, (\lambda, -) : \square, \square) && (7)\\
\mapsto &E'(\vec{V}G \otimes H')@, (e, \downarrow, \square, (\lambda, -) : \square, \square) && (8)\\
\mapsto &E'(\vec{V}G \otimes H')@, (e_2, \uparrow, \square, \star : \square, \square) && (9)\\
\mapsto &E'(\vec{V}(?([\![t]\!]\lambda)!) \otimes H')@, (e_2, \uparrow, \square, \star : \square, \square) && (10)\\
\mapsto &E'(\vec{V}[\![t]\!] \otimes H')@, (e_2, \uparrow, \square, \star : \square, \square) && (11)
\end{aligned}
$$

where $G = ?([\![t]\!])\lambda!$ and $a$ is the in-port of the $\lambda$-node. Going from (2) to (3) we apply the condition on $H \in T_{\kappa_1}$ and Cor. D.1.1 and thus $H'$ is a *value dag* and $H' \in T_{\kappa_1}$. Now by using Lem. D.21, we have $H' \prec^+ \vec{C}H^*$ such that $H^*$ a value tree and $\vec{C}H^* \in T_{\kappa_1}$. By using Lem. D.23, $H^* \in T_{\kappa_1}$.

Now we can use the induction hypothesis on $t$ to get $(\vec{V} \otimes H^*)[\![t]\!] \in T_{\kappa_2}$. Then by Lem. D.23, we have that $(\vec{V} \otimes \vec{C}H^*)[\![t]\!] \in T_{\kappa_2}$. Thus by using Prop. D.18, we have $(\vec{V} \otimes H')][\![t]\!] \in T_{\kappa_2}$. Therefore by using Lem. D.25, we have that $(\vec{V}G \otimes H)@ \in T_{\kappa_2}$.

For (c), it is straightforward as the resulting graph does not change and thus the condition is proved in (b)

**Inductive Step** $\Gamma \vdash t\ u : \tau$. In this case, we need to show that $\vec{V}G \in T_\kappa$ where $G = \vec{C_2}([\![t]\!] \otimes [\![u]\!])@$.

First by Lem. D.20 we have that $\vec{V}\vec{C_2}([\![t]\!] \otimes [\![u]\!])@ \prec^+ \vec{C_2}(\vec{V}[\![t]\!] \otimes \vec{V}[\![u]\!])@$. Therefore by Prop D.18, it suffices to show that $\vec{C}(\vec{V}[\![t]\!] \otimes \vec{V}[\![u]\!])@ \in T_\kappa$. Since by induction hypothesis on $t$ and $u$, we have that $(\vec{V}[\![t]\!] \otimes \vec{V}[\![u]\!])@ \in T_\kappa$ and then by Lem. D.23, we have that $\vec{C}(\vec{V}[\![t]\!] \otimes \vec{V}[\![u]\!])@ \in T_\kappa$.

**Base Case ($\Gamma \vdash op : \tau$).** We are going to show the case for $op = \$$ and $op = \text{link}$.

**Sub-Case** $\Gamma \vdash \$ : Int \to Int \to Int$. In this case, since $\vec{V}$ is empty, we need to show that $G \in T_{!(!Int \to !Int \to !Int)}$. This graph indeed terminates and result in the same graph, therefore we only need to prove that for any $H_1 : \{e_1 : !Int\} \to \Gamma'$, such that $H_1 \in T_{!Int}$, $(G \otimes H_1)@ \in T_{!(Int \to Int)}$.

To prove the first condition, assuming a dataflow environment $E$, we would have a sequence $\mathit{Init}(E(G \otimes H_1)@, e_2) \mapsto^* \mathit{Final}(E'H'_1G, e_2, (\lambda, -))$ by using the condition of $H_1$ such that $H'_1 \in T_{!Int}$ where $e_2$ is the in-port of the bottom @-node. Therefore it terminates and we then need to prove that $H'_1G \in T_{!(Int \to Int)}$. This graph indeed terminates and result in the same graph. Therefore we only need to prove that for any $H_2 : \{e_3 : !Int\} \to \Gamma''$ such that $H_2 \in T_{!Int}$, $(H'_1G \otimes H_2)@ \in T_{!Int}$. Since it is of ground type therefore we only need to prove that it terminates. Assuming a dataflow environment $E$ we would have a sequence from $\mathit{Init}(E(H'_1G \otimes H_2)@, e_3) \mapsto^* (E'(H'_1 \otimes H'_2)\$, e_3, \downarrow, \$, X : S, \emptyset)$. Depending on the value of $X$, it would transit to either one of the followings:

$$E'(H'_1 \otimes H'_2)\$, (e, \downarrow, \square, (n, g) : \square, \square) \tag{12}$$

$$E'(H'_1C \otimes H'_2C)(n)!, (e, \downarrow, \square, (n, -) : \square, \square) \tag{13}$$

Therefore we have proved that for any dataflow environment, the graph indeed terminates. Since in the first case, $e_3$ will be connected to a dataflow environment, thus we take the resulting term graph to be the empty graph and thus for any dataflow environment it terminates. As for the second case a constant always satisfies the logical predicate.

To prove the second condition $(G \otimes H_1)@ \in T_{!(Int \to Int)}$, we need to prove that for any $H_2 : \{e_3 : !Int\} \to \Gamma''$ such that $H_2 \in T_{!Int}$, $((G \otimes H_1)@ \otimes H_2)@ \in T_{!Int}$. The proof is similar to the case above.

**Sub-Cases** $\Gamma \vdash \mathrm{link} : \mathit{Cell} \to \mathit{Int} \to \mathit{Unit}$ **and** $\Gamma \vdash \mathrm{assign} : \mathit{Cell} \to \mathit{Int} \to \mathit{Unit}$. These cases are similar to the \$-case but also needs *validity* (Lem. D.2) to prove that the cell being update exists.

**Sub-Cases** $\Gamma \vdash \mathrm{root} : \mathit{Cell} \to \mathit{Int} \to \mathit{Unit}$. This case is similar to the link-case but we also need to argue that when the token comes back to perform the $r$-rewrite, it would always be connected to a dataflow environment as such it will terminate and that the resulting graph satisfies the logical predicate.

**Inductive Step** $\Gamma \vdash \mathrm{if}\ t\ \mathrm{then}\ t_1\ \mathrm{else}\ t_2 : \gamma$. Since only ground types are allowed in the branches, this case follows the same technique shown in the previous case. □

**Proposition D.27** (Termination). *For any closed well-typed term* $- \vdash t : \tau$ *which is recursion-free,* $\mathit{Init}([\![- \vdash t : \tau]\!], r) \mapsto^* \mathit{Final}(G, r, X)$ *for some* $G, X$ *where* $X \in \mathit{Ans}_r$ *and* $r$ *is the root of* $[\![- \vdash t : \tau]\!]$.

*Proof.* This is just a consequence of the fundamental lemma (Lem. D.26). □

### D.10 Type Soundness

**Proposition D.28** (Interface preservation). *For any valid state* $(G, \delta)$ *where* $G : \Lambda \to \emptyset$, *all transitions send* $G$ *to another graph* $G'$ *with the same interface, i.e.*

$$(G : \Lambda \to \emptyset) \wedge (G, \delta) \mapsto (G', \delta') \implies G' : \Lambda \to \emptyset.$$

*Proof.* All the pass transitions indeed satisfy the statement as they do not change the shape of the graph at all. As for rewrite transitions, the proof is done by inspecting them one-by-one. □

**Corollary D.28.1.** *For any valid state* $(G, \delta)$ *where* $G : \Lambda \to \emptyset$*, if* $(G, \delta) \mapsto^* (G', \delta')$ *then* $G' : \Lambda \to \emptyset$*.*

**Lemma D.29.** *For any valid state* $\pi$*, either*

1. $\pi$ *is a answer state, or*
2. *there exists some* $\pi'$ *such that* $\pi \mapsto \pi'$*.*

*Proof.* This is proved by case analysis on all possible valid states. □

**Proposition D.30** (Progress)**.** *For any closed well-typed term* $\vdash t : \tau$*, if Init*$([\![\vdash t : \tau]\!], r) \mapsto^n (G, \delta)$ *for some* $G, \delta$ *where* $r$ *is the root of* $[\![\vdash t : \tau]\!]$*, then either* $(G, \delta)$ *is a final state or there exists another state* $(G', \delta')$ *such that* $(G, \delta) \mapsto (G', \delta')$*.*

*Proof.* Since the initial state is valid therefore this is simply the consequence of Cor. D.2.1 and Lem. D.29. □

**Theorem D.31** (Type soundness)**.** *For any closed well-typed term* $- \vdash t : \tau$*, let* $G : \{r : [\![\tau]\!]\} \to \emptyset = [\![- \vdash t : \tau]\!]$ *then*

1. *if* $t$ *is recursion-free then the execution from Init*$(G, r)$ *will terminate,*
2. *if* $t$ *is with recursion then the execution from Init*$(G, r)$ *will not crash.*

*Proof.* This is just a consequence of Cor. D.28.1, Prop. D.27 and Prop. D.30. □

## About the authors

**Steven W.T. Cheung** is a PhD student at the University of Birmingham. Contact him at wtc488@cs.bham.ac.uk.

**Dan R. Ghica** is a Professor of Semantics of Programming Languages at the University of Birmingham. Contact him at d.r.ghica@cs.bham.ac.uk.

**Koko Muroya** is an Assistant Professor at RIMS, Kyoto University. Contact her at kmuroya@kurims.kyoto-u.ac.jp.